\documentclass[lettersize,journal]{IEEEtran}
\usepackage{amsmath,amsfonts,amsthm}
\usepackage[ruled,vlined,linesnumbered]{algorithm2e}
\usepackage{array}
\usepackage[caption=false,font=normalsize,labelfont=sf,textfont=sf]{subfig}
\usepackage{textcomp}
\usepackage{stfloats}
\usepackage{url}
\usepackage{verbatim}
\usepackage{graphicx}
\usepackage{cite}
\usepackage[short]{optidef}
\usepackage{multicol}
\usepackage{comment}
\usepackage{xcolor}
\usepackage{caption}
\usepackage{wrapfig}
\usepackage{longtable} 
\usepackage{booktabs}
\usepackage{adjustbox}
\usepackage{tikz}
\usetikzlibrary{fit,calc}
\newcommand*{\tikzmk}[1]{\tikz[remember picture,overlay,] \node (#1) {};\ignorespaces}
\newcommand{\boxit}[1]{\tikz[remember picture,overlay]{\node[yshift=3pt,fill=#1,opacity=.14,fit={(A)($(B)+(.8\linewidth,.85\baselineskip)$)}] {};}\ignorespaces}

\usepackage[hidelinks]{hyperref}
\usepackage{url}
\definecolor{Burgundy}{RGB}{144,0,32}

\definecolor{col_rollout}{RGB}{174,65,50}
\definecolor{col_ini}{RGB}{180,101,4}
\definecolor{col_solve}{RGB}{16,115,158}
\definecolor{col_eval}{RGB}{14,128,136}
\definecolor{col_shift}{RGB}{86,81,126}

\newtheorem{assumption}{Assumption}

\newtheorem{theorem}{Theorem}
\newtheorem{corollary}[theorem]{Corollary}
\newtheorem{lemma}[theorem]{Lemma}

\usepackage{acronym}

\newacro{DM}{decision making}
\newacro{AD}{automated driving}

\newacro{NLP}{nonlinear program}
\newacro{MPC}{nonlinear model predictive control}
\newacro{NMPC}{nonlinear model predictive control}
\newacro{MPPI}{model predictive path integral control}
\newacro{QP}{quadratic program}
\newacro{MIQP}{mixed-integer quadratic program}
\newacro{MILP}{mixed-integer linear program}
\newacro{MINLP}{mixed-integer nonlinear program}
\newacro{MI}{mixed-integer}
\newacro{MIP}{mixed-integer programming}
\newacro{BB}{branch-and-bound}
\newacro{SQP}{sequential quadratic programming}
\newacro{RNN}{recurrent neural network}
\newacro{OCP}{optimal control problem}
\newacro{LQR}{linear quadratic regulator}
\newacro{iLQR}{iterative linear quadratic regulator}
\newacro{SLQ}{sequientail linear quadratic programming}
\newacro{DDP}{differential dynamic programming}
\newacro{MCP}{mixed complementarity problem}
\newacro{IP}{interior point}
\newacro{ADMM}{alternating direction method of multipliers }
\newacro{RTI}{real-time iteration}
\newacro{MDP}{Markov decision process}
\newacro{SOC}{stochastic optimal control}

\newacro{RL}{reinforcement learning}
\newacro{DP}{dynamic programming}
\newacro{LSTM}{long short-term memory}
\newacro{NN}{neural network}
\newacro{PPO}{proximal policy optimization}
\newacro{PG}{policy gradient}
\newacro{TD}{temporal difference}
\newacro{SAC}{soft actor critic}
\newacro{PI}{policy iteration}
\newacro{SARSA}{state action reward state action}
\newacro{IL}{imitation learning}
\newacro{MLE}{maximum likelihood}
\newacro{SV}{surrounding vehicle}
\newacro{EV}{ego vehicle}

\newacro{FF}{feed forward network}
\newacro{DNN}{deep neural network}

\newacro{UGV}{unmanned ground vehicle}

\newcommand{\snowhill}{\emph{snow hill} environment}

\newacro{AC4MPC}{Actor Critic for Nonlinear Model Predictive Control}

\newcommand{\ub}[1]{\overline{#1}}
\newcommand{\lb}[1]{\underline{#1}}

\newcommand{\R}{\mathbb{R}}

\DeclareMathSymbol{\shortminus}{\mathbin}{AMSa}{"39}

\newcommand{\eval}{\mathrm{ac4eval}}

\newcommand{\shift}{\mathrm{shift}}

\newcommand{\intSt}[1]{\mathbb{N}_{#1}}
\newcommand{\intSet}[2]{\mathbb{N}_{[{#1},{#2}]}}

\DeclareMathOperator*{\E}{\mathbb{E}}

\begin{document}

\title{AC4MPC: Actor-Critic Reinforcement Learning for Nonlinear Model Predictive Control}

\author{Rudolf Reiter, Andrea Ghezzi, Katrin Baumgärtner, Jasper Hoffmann, Robert D. McAllister, Moritz Diehl
\thanks{This research was supported by DFG via Research Unit FOR 2401 and project 424107692 and by the EU via ELO-X 953348. \emph{Corresponding Author: Rudolf Reiter}}
\thanks{Rudolf Reiter, Andrea Ghezzi and Katrin Baumgärtner are with the Department of Microsystems Engineering (IMTEK), University of Freiburg, 79110 Freiburg, Germany (e-mail: \{rudolf.reiter, andrea.ghezzi, katrin.baumgaertner\}@imtek.uni-freiburg.de).}
\thanks{Jasper Hoffmann is with the Department of Computer Science, University of Freiburg, 79110 Freiburg, Germany (e-mail: hofmaja@informatik.uni-freiburg.de).}
\thanks{Robert D. McAllister is with the Delft Center for Systems and Control, Delft University of Technology, 2628 Delft, The Netherlands (e-mail: r.d.mcallister@tudelft.nl).}
\thanks{Moritz Diehl is with the Department of Microsystems Engineering and the Department of Mathematics, University of Freiburg, 79110 Freiburg, Germany (e-mail: moritz.diehl@imtek.uni-freiburg.de).}}

\markboth{Journal of \LaTeX\ Class Files,~Vol.~14, No.~8, August~2021}%
{Shell \MakeLowercase{\textit{et al.}}: A Sample Article Using IEEEtran.cls for IEEE Journals}

\IEEEpubid{0000--0000/00\$00.00~\copyright~2021 IEEE}

\maketitle

\begin{abstract}
\Ac{MPC} and \ac{RL} are two powerful control strategies with, arguably, complementary advantages. 
In this work, we show how actor-critic \ac{RL} techniques can be leveraged to improve the performance of \ac{MPC}. The \ac{RL} critic is used as an approximation of the optimal value function, and an actor roll-out provides an initial guess for primal variables of the \ac{MPC}. 
A parallel control architecture is proposed where each \ac{MPC} instance is solved twice for different initial guesses. 
Besides the actor roll-out initialization, a shifted initialization from the previous solution is used.
Thereafter, the actor and the critic are again used to approximately evaluate the infinite horizon cost of these trajectories.
The control actions from the lowest-cost trajectory are applied to the system at each time step.
We establish that the proposed algorithm is guaranteed to outperform the original \ac{RL} policy plus an error term that depends on the accuracy of the critic and decays with the horizon length of the \ac{MPC} formulation.
Moreover, we do not require globally optimal solutions for these guarantees to hold.
The approach is demonstrated on an illustrative toy example and an \ac{AD} overtaking scenario.
\end{abstract}

\begin{IEEEkeywords}
model predictive control, reinforcement learning, dynamic programming
\end{IEEEkeywords}
\acresetall
\section{Introduction}
\IEEEPARstart{I}{n} \ac{MPC}, an optimization problem comprising the cost of a simulated trajectory of an environment model, is solved online in each control iteration~\cite{Rawlings2017}.
The optimization framework provides an intuitive and effective way for devising nonlinear controllers suitable for a wide range of applications~\cite{Schwenzer2021}. For instance, it allows direct optimization over the desired cost function or specific constraints.
However, for constrained systems with fast dynamics or scarce computational resources, the computational demands associated with solving the optimization problem within the available sampling time are often prohibitive, limiting the adoption of \ac{MPC} in many applications.
A common way to decrease the computation time is to approximate the value function~\cite{Zhong2013} and a control invariant set to use as terminal cost and constraint, respectively~\cite{Rawlings2017}.
Hence, if the adopted approximations are sufficiently accurate, the~\ac{MPC} horizon length can be reduced. A shorter horizon length reduces the number of decision variables and, therefore, computation time. 

Another challenge with nonlinear \ac{MPC} arises if the optimization problem is solved by direct methods, which formulate it as a \ac{NLP}~\cite{Bock1984}.
These methods require an initial guess close to the, preferably global or sufficiently good, local optimal solution to avoid getting stuck in a bad solution. Sufficiently good initial guesses reduce the number of iterations required for the optimization algorithm to converge.

\Ac{RL}, in contrast, is a collection of algorithms that aim at solving a \ac{MDP}, which is equivalent to a stochastic~\ac{OCP}~\cite{Zanon2022}.
By principles of \ac{DP}~\cite{Bellman1966, Bertsekas1995}, an optimal policy, also referred to as \emph{actor}, and an optimal value function, also referred to as \emph{critic}, are approximated by parameterized functions and trained during interaction with the environment.
Intrinsic to all \ac{RL} algorithms is the goal of obtaining globally optimal policies and value functions via interactions with the environment.
\Ac{RL} usually obtains policies with a low accuracy but close to global solutions, in contrast to \ac{MPC}, which finds high-accuracy local solutions. 
This behavior of \ac{RL} algorithms is due to the potentially high-dimensional state and action spaces, limited number of samples, and limited expressiveness of the \acp{NN}.
However, \acp{NN} policies usually have a low online computation time.
\IEEEpubidadjcol

Remarkably, these properties are nearly orthogonal to those of \ac{MPC}~\cite{Gorges2017}.
The algorithm proposed in this paper is referred to as~\ac{AC4MPC} and combines these complementary advantages. 
\Ac{AC4MPC} aims at obtaining globally optimal policies by using trained \acp{NN} of actor-critic~\ac{RL} algorithms to construct a terminal value function and a policy roll-out to provide an initial guess for~\ac{MPC}.
To obtain fast online computation times, we propose a framework to augment the \ac{RTI} scheme~\cite{Diehl2005} with a parallel optimizer and evaluate whether this parallel solution exhibits a lower cost.
Evaluating the parallel solutions for their predicted performance at each iteration is nontrivial since it may be an intermediate solution of the \ac{RTI} scheme.
Again, we utilize the actor and critic networks for an auxiliary evaluation control law, terminal roll-out, and terminal value function to rank the trajectories among their predicted cost and select the lowest-cost control for each iteration.

\subsection{Related Work}
Due to their complementary advantages, \ac{MPC} and \ac{RL} have been previously combined in various ways. To promote sample efficiency and safety, \cite{Amos2018,Gros2020,Gros2021a,Romero2023,Reiter2023} use MPC together with \acp{NN} within the RL policy, and~\cite{Ghezzi2023} uses an MPC formulation within the RL critic. These methods do not address the difficulties of MPC warm-starts and terminal cost approximations.

The presented approach flips the paradigm of~\cite{Levine2013,Lowrey2019,Wang2023}, which uses \ac{MPC} as an expert to warm-start the training of an actor-critic \ac{RL} algorithm. In fact, we assume a well-trained actor-critic \ac{RL} to warm start the online optimization of an \ac{MPC} to improve the overall performance.

Allocating warm starts by external modules such as~\ac{NN} and using approximations of the terminal value function have been studied in various works.
For instance, in~\cite{Klauco2019}, warm-starts for the active set are used, \cite{Sambharya2023} warm-starts \acp{QP} and \cite{Masti2019,Marcucci2021} use trained \acp{NN} to warm-start mixed-integer solvers.
The authors in~\cite{Mansard2018,Qu2023,Chen2022} use a trained \ac{NN} to warm-start \ac{MPC}.
However, using external warm-starts in each iteration may conflict with the \ac{RTI} scheme~\cite{Diehl2005}.
The authors in~\cite{Shen2023} use \ac{RL} on a coarse discrete state space to provide approximate motion plans for multiple vehicles that are tracked thereafter by distributed MPC.

The authors in~\cite{Abdufattokhov2021} approximate the optimal value function for~\acp{MDP} related to regulation problems where a quadratic terminal value function is obtained by supervised learning. The authors in~\cite{Zhong2013, Beckenbach2020,Beckenbach2022,Moreno2023} use approximate \ac{DP} or Q-learning, respectively, to approximate the value function for \ac{MPC} and~\cite{Deits2019} use combinatorial optimization solver evaluations to approximate the value function related to an mixed-integer problem. The authors in~\cite{Beckenbach2020,Beckenbach2022} provide stability but also require a certain structure of the cost function. Similar to the proposed approach,~\cite{Karnchanachari2020} learns a value function as part of an \ac{MPC} policy within an actor critic method. 
The authors in \cite{Karnchanachari2020} do not make use of the actor nor parallel computations or evaluations, yet, they state relevant practical considerations when using \ac{SQP} with \acp{NN}. Using the \ac{MPC} policy as the actual actor within \ac{RL} may be an additional extension to the presented framework. 

If the system can be stabilized around a reference, also a stabilizing control law can be used to approximate the terminal value function within MPC. This was shown in~\cite{Nicolao1998,Diehl2003e, Diehl2004f} with a stabilizing LQR policy.

The author in~\cite{Bertsekas2005} summarizes several fundamental concepts used within this work, i.e., suboptimal control, explicit value function approximations, and roll-outs as implicit approximations. \Ac{AC4MPC} can be seen as a specific suboptimal control algorithm to approximate the optimal policy and value function, respectively.

\subsection{Contribution}
The contributions of this work are the following:
\begin{itemize}
	\item derivation of a control strategy, namely AC4MPC, that combines \ac{MPC} and \ac{RL} to improve the overall performance,
	\item theoretical justification of the closed-loop performance improvement that does not rely on globally optimal solutions of the \ac{AC4MPC} optimization problem and is therefore consistent with \ac{RTI} schemes,
	\item derivation of an real-time capable algorithm based on AC4MPC and RTI, referred to as AC4MPC-RTI,
	\item evaluation of AC4MPC-RTI on a realistic autonomous driving simulation.
\end{itemize}

\subsection{Outline}
The remainder of the paper is structured as follows. In Sect.~\ref{sec:preliminaries}, we state the main concepts used within this paper, which are NMPC and actor-critic RL. In Sect.~\ref{sec:ac4mpc} the main algorithm, \ac{AC4MPC}, is introduced and its theoretical properties are derived.
The method is furthermore adapted to yield a real-time capable version, \ac{AC4MPC}-RTI, in Sect.~\ref{sec:ac4mpc_parallel}.
In Sect.~\ref{sec:experiments}, the performance on an illustrative example, and a more realistic \ac{AD} example are evaluated.
We conclude and discuss the paper in Sect.~\ref{sec:discussion}.

\section{Preliminaries}
\label{sec:preliminaries}
This section introduces the problem setup and important concepts from both \ac{RL} and \ac{MPC}.

Indices~$k$ are used for predictions, i.e., roll-outs, at a current time step~$j$.
We refer to~$\mathbb{N}=\{0,1,\ldots\}$ as the natural numbers including zero. 
We use the definitions~$\intSt{N}=\{x \in \mathbb{N} \; | \; x\leq N\}$ and $\intSet{n}{N}=\{x \in \intSt{N}\; | \; n\leq x\}$.
The vector of ``all ones'' is~$\textbf{1}$ with suitable dimensions. The Huber function is defined as
\begin{equation*}
	H_{\delta}(x) = \begin{cases} 
		\frac{1}{2}x^2 & \text{if } |x| \leq \delta \\
		\delta(|x| - \frac{1}{2}\delta) & \text{otherwise}
	\end{cases}
\end{equation*}
The terminology of the control systems literature is used with some slight modifications.
We consider the nominal case, i.e., the system is assumed to be deterministic as opposed to the stochastic~\ac{RL} environment.

The state~$s\in\mathbb{S}\subseteq\R^{n_s}$ and the control~$u\in\mathbb{U}\subseteq\R^{n_u}$ are related to the dynamic discrete-time Markovian environment with the transition function
\begin{equation*}
    s_{j+1}=F(s_j,u_j), \qquad F:\mathbb{S}\times\mathbb{U}\rightarrow\mathbb{S}.
\end{equation*}
Note that $\mathbb{S}$ is the domain/range of the state space, not a desired state constraint.
The objective is formulated in terms of minimizing a non-negative cost function~$c(s,u):\R^{n_s}\times\R^{n_u}\rightarrow\R_{\geq 0}$, rather than maximizing a reward.
With a discount factor~$\gamma\in(0,1]$, the value function of a control law or policy~$\pi(s):\mathbb{S}\rightarrow\mathbb{U}$ is
\begin{align}
	\begin{split}
	\label{eq:general_expected_return}
	J_\pi(s)  \coloneqq&  \sum_{k=0}^{\infty} \gamma^k c(s_k,u_k) ,\\
	&  s_0 = s,\; s_{k+1}= F(s_k,u_k),\; u_{k} = \pi(s_k),
	\end{split}
\end{align}
and the \ac{OCP} that defines the optimal cost~$J^*(s)$ for a given state~$s$ can be stated by
\begin{equation*}
    J^*(s) := \min_{\pi}J_{\pi}(s),
\end{equation*}
and the optimal policy~$\pi^*$ is defined such that~$\pi^*(s)=\arg \min_\pi J_{\pi}(s)$ for all~$s\in\mathbb{S}$.
The optimal Q-function is directly related to the optimal value function by
\begin{equation}
	\label{eq:Q_fun}
	Q^*(s,u) := c(s,u) + \gamma J^*\big(F(s,u)\big).
\end{equation}

We include constraints within the cost function, i.e., we are rewriting hard constraints via L1-penalties.
Particularly, a priority over a nominal cost~$c_0(s, u)$ is given to satisfying equality constraints~$g(s,u)=0$ with $g(s,u):\R^{n_s}\times\R^{n_u}\rightarrow\R^{n_g}$ and constraints~$h(s,u)\geq 0$ with $h(s,u):\R^{n_s}\times\R^{n_u}\rightarrow\R^{n_h}$
within the cost formulation
\begin{align}
	\label{eq:cost}
	\begin{split}
		c(s_k, u_k)=~&
		c_0(s_k,u_k) \\
		&+ w_g^\top |g(s_k, u_k)|+ w_h^\top \min(h(s_k, u_k),0).
	\end{split}
\end{align}
For sufficiently large weights~$w_g\in\R^{n_g}$ and~$w_h\in\R^{n_h}$, the optimal solution is equivalent to the solution of the constrained problem~\cite{Byrd2008}.

\subsection{Reinforcement Learning}
\label{sec:rl}
This work is based on actor-critic \ac{RL} algorithms to approximate~$ \pi^*(s)$,~$J^*(s)$, and/or~$Q^*(s,u)$ with parameterized functions~$\hat{\pi}(s)$,~$\hat{J}(s)$, and/or $\hat{Q}(s,u)$.
Within \ac{RL} one can distinguish between on-policy methods, such as~\ac{PPO}~\cite{Schulman2017}, that collect samples with the currently learned policy~$\hat{\pi}$ before each update, and off-policy methods, such as \ac{SAC}~\cite{Haarnoja2018b}, that use data generated from policies unrelated to the currently learned policy.
In the following, we utilize both actor-critic policy types, i.e., \ac{SAC} and \ac{PPO}. The value function obtained by \ac{SAC} is typical of the type~$\hat{Q}(s,u)$, and from \ac{PPO} it is~$\hat{J}(s)$. 

In the tabular setting, comprising discrete states and controls without function approximation, the convergence of~$\hat{\pi}(s)$,~$\hat{J}(s)$, and/or $\hat{Q}(s,u)$ towards their optimal counterparts~$ \pi^*(s)$,~$J^*(s)$, and/or~$Q^*(s,u)$ can be shown for both \ac{SAC}~\cite{Haarnoja2018} and \ac{PPO}~\cite{Kuba2022}.
For continuous state and control spaces, approximation error bounds are often restricted to linear function approximation, excluding non-linear functions like neural networks~\cite{Sutton2018}.
However, in practice, the estimates often converge towards the optimal policy~$\pi^*(s)$ and optimal value functions~$J^*(s)$ or $Q^*(s, u)$.

For more details on the policy, we refer to Appendix~\ref{sec:app_rl}.

\subsection{Nonlinear Model Predictive Control}
\label{sec:nmpc}

\ac{MPC} approximates the infinite horizon cost function in~\eqref{eq:general_expected_return} via a finite horizon~$N\in\mathbb{N}$ and a terminal cost~$V_f:\mathbb{S}\rightarrow\R_{\geq 0}$, stated as
\begin{equation}
	\label{eq:mpc_cost}
    V_N(s,\mathbf{u}) := \sum_{k=0}^{N-1}\gamma^k c(s_k, u_k) + \gamma^N V_f(s_N)
\end{equation}
in which $\mathbf{u}=(u_0,u_1,\dots,u_{N-1})$, $s_{k+1}=F(s_k,u_k)$, and $s_0=s$.
The terminal cost~$V_f(s)$ is typically designed to approximate the value function~$J_{\pi}(s)$ for a policy~$\pi(s)$ that asymptotically stabilizes the nominal system or achieves a more general performance objective. The \ac{MPC} optimization problem is then
\begin{equation}\label{eq:MPC_compact}
    V_N^0(s) := \min_{\mathbf{u}\in\mathbb{U}^N}V_N(s,\mathbf{u}).
\end{equation}
In Section~\ref{sec:ac4mpc}, we leverage the concept of terminal costs and associated theoretical results to devise the proposed method.

In order to compute the solution of~\eqref{eq:MPC_compact} we adopt a direct approach, specifically direct multiple shooting~\cite{Bock1984}, yielding the following problem formulation:
\begin{align}
	\begin{split}
		\label{eq:MPC_general}
		\min_{\mathbf{s}, \mathbf{u}}&\sum_{k=0}^{N-1}\gamma^k c(s_k, u_k) + \gamma^N V_f(s_N) \\
		&\text{s.t.}\;
		\begin{cases}
			s_0=s_j, \\
			s_{k+1}= F(s_k,u_k), \\
            u_k\in\mathbb{U},\;k\in\mathbb{N}_{N-1}.
		\end{cases}
	\end{split}
\end{align}
Let $\mathbf{s} = (s_0, \dots, s_N)$ be the vector that collects the state along the prediction horizon. We include~$\mathbf{s}$ among the optimization variables and a continuity condition for the system dynamic at each step of the control horizon.
Enlarging the dimension of the NLP with the variables in $\mathbf{s}$ makes~\eqref{eq:MPC_general} sparse and structured. These properties enhance numerical stability and improve convergence.
A favorable numerical method for solving~\eqref{eq:MPC_general} is \ac{SQP}~\cite{Nocedal1999}. It allows for effective warm-starting of the primal variables~$\textbf{s}$ and $\textbf{u}$.
Iteratively converging schemes, such as the \ac{RTI} scheme~\cite{Diehl2005}, can deal with fast sampling times and constrained memory of embedded devices.

Specifically, \ac{SQP} attains the solution of the given \ac{NLP} by iteratively solving \acp{QP} obtained by linearizing the nonlinear constraints in~\eqref{eq:MPC_general} and computing a quadratic approximation of the, potentially nonlinear, cost function.
Thus, the convergence of an \ac{SQP} algorithm to a minimizer of the NLP~\eqref{eq:MPC_general} requires the solution of potentially several \acp{QP}.

One can mitigate this burden by adopting the \ac{RTI} scheme, which performs only one, or in general~$M$, \ac{SQP} iterations per sampling time.
Intuitively, the convergence towards the minimizer of~\eqref{eq:MPC_general} takes place over consecutive time steps.
In every closed-loop iteration, the previous solution is shifted to provide the initial guess for the new OCP.
Note that within \ac{RTI}, we may apply a control action to the system that stems from an \ac{SQP} iteration which is not yet fully converged to the optimum of~\eqref{eq:MPC_general}.
Hence, the \ac{RTI} solution may violate the nonlinear constraint of~\eqref{eq:MPC_general}, e.g., the resulting trajectory may not be dynamically feasible.
This observation will be particularly important in Section~\ref{sec:ac4mpc_eval} in order to evaluate the cost of an infeasible trajectory.

Notice that solving~\eqref{eq:MPC_general} in each step to the global optimum, using a perfectly estimated value function~$V_f(s)\equiv J^*(s) $ and applying the first control, yields the optimal policy~$\pi^*(s)$.
This follows directly from the definition~\eqref{eq:general_expected_return} and~\eqref{eq:Q_fun}.
However, obtaining the global optimizer~\eqref{eq:MPC_general} is, in general, intractable.
Moreover, nonlinear optimization solvers converge to local optima depending on the solver initialization. 

\section{Actor and Critic Models for Nonlinear Model Predictive Control}
\label{sec:ac4mpc}
In general, none of the optimal functions~$ \pi^*(s)$,~$J^*(s)$, or~$Q^*(s,u)$ are available. Moreover, solving nonlinear (nonconvex) MPC optimization problems to guaranteed global optimality is often impossible in online applications. Therefore, we instead consider a suboptimal algorithm, referred to as \ac{AC4MPC}, that does not assume that an optimal solution is obtained. 
The algorithm, which is this paper's main contribution, is summarized in Alg.~\ref{alg:ac4mpc}.
\Ac{AC4MPC} uses an actor model~$\hat{\pi}(s)$ and a critic model~$\hat{J}(s)$ or~$\hat{Q}(s,u)$ to improve the performance of \ac{MPC}.
The trained actor and critic~\acp{NN} are obtained by methods described in Sect.~\ref{sec:rl}. 

\subsection{Basic AC4MPC Algorithm Description}
In \ac{AC4MPC}, the terminal cost for the standard \ac{MPC} formulation is defined by either the approximate value function~$\hat{J}(s)$, e.g., obtained by \ac{PPO}, or $Q$-value function~$\hat{Q}(s,\hat{\pi}(s))$, e.g., obtained by \ac{SAC}. Since these estimated value functions are not exact, including an additional rollout of the actor~$\hat{\pi}(s)$ can improve the estimate of the true value function for this actor policy \cite{Bertsekas2005}. For a rollout of $R\in\mathbb{N}$ and estimated value function~$\hat{J}(\cdot)$, we define the terminal cost as
\begin{equation}\label{eq:V_f}
    V_f(s) := \sum_{i=0}^{R-1} \gamma^ic(s_k,\hat{\pi}(s_k)) + \gamma^R\hat{J}(s_R)
\end{equation}
in which $s_{k+1}=F(s_k,\hat{\pi}(s_k))$ and $s_{0}=s$. For an estimated $Q$-value function, we simply replace $\hat{J}(s)$ by $\leftarrow $$\hat{Q}(s,\hat{\pi}(s))$. This rollout aims to better approximate the value function for the actor $\hat{\pi}(s)$. With this terminal cost, the \ac{MPC} objective function becomes
\begin{multline*}
    V_N(s,\mathbf{u}) = \sum_{k=0}^{N-1}\gamma^k c(s_k, u_k) \\ + \sum_{k=N-1}^{N+R-1}\gamma^k c(s_k, \hat{\pi}(s_k)) + \gamma^{N+R} \hat{J}(s_{N+R})
\end{multline*}
in which
\begin{equation*}
    s_{k+1} = \begin{cases}
        F(s_k,u_k) & k\in\mathbb{N}_{[0,N-1]} \\
        F(s_k,\hat{\pi}(s_k)) & k\in\mathbb{N}_{[N,N+R-1]}
    \end{cases}
\end{equation*}
Thus, the first $N$ inputs $u_k$ are free variables, while the following $R$ inputs are fixed by the actor $\hat{\pi}(\cdot)$. 

In addition to the rollout, the actor~$\hat{\pi}(s)$ provides an initial trajectory of states and controls for the \ac{AC4MPC} optimization problem.
Specifically, we define the simulated state and input trajectory from an initial state~$s\in\mathbb{S}$ as $\hat{\Phi}(s;\hat{\pi}(\cdot))=(\hat{s}_0,\hat{s}_1,\dots,\hat{s}_N)$ and $\hat{\Psi}(s;\hat{\pi}(\cdot))=(\hat{u}_0,\hat{u}_1,\dots,\hat{u}_{N-1})$ in which
\begin{equation}\label{eq:cl_pi}
	\hat{s}_{k+1}=F(\hat{s}_{k},\hat{u}_k), \quad \hat{u}_k=\hat{\pi}(\hat{s}_k), \quad \hat{s}_0=s.
\end{equation}

After the first initialization, the subsequent initial guess is obtained by shifting the most recent iterate and using the actor~$\hat{\pi}(s)$ to provide an initial guess only for the very last control.
Let $\mathbf{u}=(u_0,u_1,\dots,u_{N-1})$ denote the input trajectory computed by \ac{AC4MPC} for the current state~$s=s_0$. Then, we define a shifted input trajectory at the subsequent time step as
\begin{equation}
\label{eq:control_shift_actor}
	\tilde{\mathbf{u}}^+ = \zeta(s,\mathbf{u};\hat{\pi}(\cdot)) := \Big(u_1,\dots,u_{N-1},\hat{\pi}(s_N)\Big).
\end{equation}

The \ac{AC4MPC} algorithm then selects a better policy (no worse) than $\tilde{\mathbf{u}}$ and $\hat{\Psi}(s)$, i.e., produces a lower value of $V_N(\cdot)$.
Thus, the \ac{AC4MPC} algorithm implicitly defines a function~$K_N:\mathbb{S}\times\mathbb{U}^{N}\rightarrow \mathbb{U}^N$, which satisfies
\begin{equation}\label{eq:K_N}
	K_N(s,\tilde{\mathbf{u}}) \in \left\{\mathbf{u}\in\mathbb{U}^N \ \middle| \ \begin{matrix} V_N(s,\mathbf{u}) \leq V_N(s,\tilde{\mathbf{u}}),\\
    V_N(s,\mathbf{u}) \leq V_N(s,\hat{\Psi}(s;\hat{\pi}(\cdot)))\end{matrix}\right\}.
\end{equation}
Note that the global optimum of the \ac{AC4MPC} optimization problem satisfies the requirements of $K_N(s,\tilde{\mathbf{u}})$. The control policy~$\kappa_N:\mathbb{S}\times\mathbb{U}^N\rightarrow \mathbb{U}$ defined by \ac{AC4MPC} is the first input in the trajectory defined by $K_N(s,\tilde{\mathbf{u}})$, i.e.,
\begin{equation*}
	\kappa_N(s,\tilde{\mathbf{u}}) := u_0 \quad \textrm{with} \quad (u_0,u_1,\dots,u_{N-1})=K_N(x,\mathbf{u}).
\end{equation*}
With this control policy, we obtain the closed-loop system-optimizer dynamics
\begin{equation}\label{eq:cl}
\begin{split}
	s_{j+1} & = F(s_j,u_j), \quad u_j=\kappa_N(s_j,\tilde{\mathbf{u}}_j), \\
    \tilde{\mathbf{u}}_{j+1} & = \zeta(s_j,K_N(s_j,\tilde{\mathbf{u}}_j);\hat{\pi}(\cdot)).
\end{split}
\end{equation}
Note that both the state~$s_j$ and the initial guess~$\tilde{\mathbf{u}}_j$ evolve according to autonomous dynamics defined by \ac{AC4MPC}. 
Alg.~\ref{alg:ac4mpc} provides a simple example of an \ac{AC4MPC} algorithm that satisfies the requirements in~\eqref{eq:K_N}. 
Different conceptual parts are highlighted in color and aligned with the associated parts in the following chapters, i.e., a policy roll-out (red), the initialization of the \ac{MPC} (yellow), obtaining the solution of the \ac{MPC} (blue), evaluating different trajectories (green), and shifting and simulating the last control (purple).
Note that the solution to the \ac{MPC} problem in Alg.~\ref{alg:ac4mpc} (i.e., \textit{solve} \textsf{MPC}) does not need to be a global optimum. 

\IncMargin{1em}
\begin{algorithm}
	\SetKwData{Left}{left}
	\SetKwData{This}{this}
	\SetKwData{Up}{up}
	\SetKwData{MPC}{MPC}
	\SetKwFunction{Union}{Union}
	\SetKwFunction{FindCompress}{FindCompress}
	\SetKwInOut{Input}{input}
	\SetKwInOut{Output}{output}
	\Input{Policy $\hat{\pi}(\cdot)$, value function $\hat{Q}(\cdot)$ or $\hat{J}(\cdot)$}
	\BlankLine
	\emph{\MPC $\leftarrow$ formulation~\eqref{eq:MPC_general} with $V_f(\cdot)$ defined in~\eqref{eq:V_f} with $\hat{J}(\cdot)$ or $\hat{J}(\cdot)=\hat{Q}(\cdot,\hat{\pi}(\cdot))$}\;
	\For{$j\leftarrow 0$ \KwTo $\infty$}{
		\emph{$s\leftarrow$state measurement}\;
		\tikzmk{A}\emph{policy roll-out $\hat{\mathbf{u}} \leftarrow  \hat{\Psi}(s;\hat{\pi}(\cdot))$}\;\tikzmk{B}
        \boxit{col_rollout}
		\tikzmk{A}\If{$j == 0$}{
			$\tilde{\mathbf{u}}\leftarrow \hat{\mathbf{u}} $\;}
       \emph{initialize \MPC$\leftarrow \tilde{\mathbf{u}} $}\;\tikzmk{B}
        \boxit{col_ini}
		\tikzmk{A}\emph{$\mathbf{u}\gets$solve \MPC}\;\tikzmk{B}
        \boxit{col_solve}
  		\tikzmk{A}\If{$V_N(s,\hat{\mathbf{u}})\leq V_N(s,\mathbf{u})$}{
			\emph{$\mathbf{u}\gets\hat{\mathbf{u}}$}\;
		}\tikzmk{B}
        \boxit{col_eval}
		\emph{apply $u\gets \mathbf{u}[0]$ to the system}\;
        \tikzmk{A}
        \emph{shifting $\tilde{\mathbf{u}}\gets\zeta(s,\mathbf{u};\hat{\pi}(\cdot))$}\;\tikzmk{B}
        \boxit{col_shift}
	}
	\caption{AC4MPC}\label{alg:ac4mpc}
\end{algorithm}\DecMargin{1em}

\subsection{Cost Reduction and Performance}
\label{sec:ac4mpc_cost_reduction}
In this section, we provide some theoretical justification for the proposed algorithm. Specifically, we establish that the closed-loop performance of the proposed \ac{AC4MPC} algorithm is at least as good as the actor and critic used in the algorithm. To establish this result, we require the following assumption for the actor and critic pair. While we focus on the value function approximation~$\hat{J}(\cdot)$ and the NMPC problem in~\eqref{eq:MPC_general}, these results can be easily extended to a Q-function, e.g., we can choose $\hat{J}(s)=\hat{Q}(s,\hat{\pi}(s))$.

\begin{assumption}[Bellman Error]\label{as:critic}
	The function~$\hat{J}(s)$ is continuous and there exists~$\delta\geq 0$ such that
	\begin{equation}\label{eq:bellman}
		\lvert  \hat{J}(s) - c(s,\hat{\pi}(s)) - \gamma\hat{J}(F(s,\hat{\pi}(s))\rvert \leq \delta \quad \forall \ s\in\mathbb{S}
	\end{equation}
\end{assumption}
With this assumption, we establish a cost decrease inequality for \ac{AC4MPC}.

\begin{lemma}[Cost decrease]\label{lem:costdec}
	If Assumption~\ref{as:critic} holds, then
	\begin{align}\label{eq:costdec}
		\begin{split}
			&\gamma V_N\Big(s^+,K_N(s^+,\tilde{\mathbf{u}}^+)\Big) - V_N\Big(s,K_N(s,\tilde{\mathbf{u}})\Big)\\ & \leq - c\Big(s,\kappa_N(s,\tilde{\mathbf{u}})\Big) + \gamma^{N+R}\delta
		\end{split}
	\end{align}
	in which $s^+=F(s,\kappa_N(s,\tilde{\mathbf{u}}))$ and $\tilde{\mathbf{u}}^+=\zeta(s,K_N(s,\tilde{\mathbf{u}});\hat{\pi}(\cdot))$ for all $s\in\mathbb{R}^{n_s}$ and $\tilde{\mathbf{u}}\in\mathbb{U}^{N}$.
\end{lemma}

\begin{proof}
	For any $s\in\mathbb{S}$ and $\tilde{\mathbf{u}}\in\mathbb{U}^{N}$, let $s^+=F(s,\kappa_N(s,\tilde{\mathbf{u}}))$ and $\tilde{\mathbf{u}}^+=\zeta(s,K_N(s,\tilde{\mathbf{u}});\hat{\pi}(\cdot))$. Let $s_{k}$ denote the open-loop state at time $k\in\mathbb{N}_N$ for given $s_0=s$ and the input trajectory $\mathbf{u}=(u_0,u_1,\dots,u_{N-1})$. Let $s_{k+1}=F(s_{k},\hat{\pi}(s_{k}))$ for all $k\in\mathbb{N}_{[N,N+R]}$. From the definition of $\tilde{\mathbf{u}}^+$, we have
	\begin{multline*}
		\gamma V_N\Big(s^+,\tilde{\mathbf{u}}^+\Big) - V_N\Big(s,K_N(s,\tilde{\mathbf{u}})\Big) \\ = -c\Big(s,\kappa_N(s,\tilde{\mathbf{u}})\Big) + \gamma^{N+R+1}\hat{J}(s_{N+R+1}) + \\ \gamma^{N+R}c(s_{N},\hat{\pi}(s_{N+R})) - \gamma^{N+R}\hat{J}(s_{N+R})
	\end{multline*}
	From~\eqref{eq:bellman}, we have
	\begin{multline*}
		\gamma V_N\Big(s^+,\tilde{\mathbf{u}}^+\Big) - V_N\Big(s,K_N(s,\tilde{\mathbf{u}})\Big) \\ \leq -c\Big(s,\kappa_N(s,\tilde{\mathbf{u}})\Big) + \gamma^{N+R}\delta
	\end{multline*}
	From the definition of $K_N(\cdot)$, we have
	\begin{equation*}
		V_N\Big(s^+,K_N(s^+,\tilde{\mathbf{u}}^+)\Big) \leq V_N\Big(s^+,\tilde{\mathbf{u}}^+\Big)
	\end{equation*}
	and combining these equations gives~\eqref{eq:costdec}.
\end{proof}

If $\gamma<1$,~\eqref{eq:costdec} indicates that longer horizons~$N$ and rollouts~$R$ in \ac{AC4MPC} reduce the effect of the Bellman error in the critic~$\hat{J}(s)$. Thus, the effect of the value function $J^(s)$ is negligible for sufficiently long horizons and rollouts. At the other extreme, \ac{AC4MPC} with $N=1$ and $R=0$ is equivalent to one value iteration of the critic~$\hat{J}(s)$. These observations are, of course, consistent with results for~$\ell$-step lookahead algorithms in dynamic programming (see, e.g., \cite{Bertsekas2023}). The novel contribution of Lemma~\ref{lem:costdec} is that this property also holds for the proposed \emph{suboptimal} algorithm. 

Note that if~$c(s,u)$ is a (continuous) positive definite function with respect to the origin and $\gamma=1$, then~\eqref{eq:costdec} is equivalent to the cost decrease condition required for~$V_N(\cdot)$ to be a (practical) Lyapunov function for the closed-loop system. The stability of the origin then follows from standard assumptions about the continuity of $V_f(\cdot)$ or $V_N(\cdot)$ at the origin \cite[s. 2.4.2]{Rawlings2017}.

Given that we are also interested in systems that may have unreachable setpoints and/or cost functions that are not necessarily positive definite with respect to the origin, we instead focus the following results on the performance of the closed-loop system in terms of the stage cost~$c(s,u)$. In particular, we are interested in the closed-loop performance of the \ac{AC4MPC} algorithm defined by
\begin{equation*}
	\mathcal{J}_T(s) := \sum_{j=0}^{T-1}\gamma^jc(s_j,u_j) \quad \textrm{s.t.} \ \eqref{eq:cl},
\end{equation*}
relative to the closed-loop performance of the actor~$\hat{\pi}(s)$ defined by
\begin{equation*}
	\hat{\mathcal{J}}_T(s) := \sum_{j=0}^{T-1}\gamma^jc(\hat{s}_j,\hat{u}_j) \quad \textrm{s.t.} \ \eqref{eq:cl_pi}.
\end{equation*}

We can establish the following bound for the transient closed-loop system without any additional assumptions.

\begin{theorem}[Performance]\label{lem:performance}
	If Assumption~\ref{as:critic} holds, then
	\begin{multline}\label{eq:full_performance}
		\mathcal{J}_T(s) - \hat{\mathcal{J}}_T(s)
		\leq \\ \gamma^{N+R}\hat{J}(\hat{s}_N) + \gamma^{N+R}\delta\sum_{j=0}^{T-1}\gamma^{j} - \sum_{j=N+R}^{T-1}\gamma^{j}c(\hat{s}_j,\hat{u}_j)
	\end{multline}
	for all $s\in\mathbb{S}$ and $T\geq N+R$.
\end{theorem}

\begin{proof}
	Choose $s\in\mathbb{S}$ and rearrange~\eqref{eq:costdec} to give
	\begin{align*}
		&\gamma^{j}c\Big(s_j,\kappa_N(s_j,\tilde{\mathbf{u}}_j)\Big) \leq \gamma^{N+R+j}\delta +\\
		&\gamma^{j}V_N\Big(s_j,K_N(s_j,\tilde{\mathbf{u}}_j)\Big) - \gamma^{j+1} V_N\Big(s_{j+1},K_N(s_{j+1},\tilde{\mathbf{u}}_{j+1})\Big)
	\end{align*}
	for the closed-loop system and all $j\in\mathbb{N}$. We sum both sides of this inequality from $j=0$ to $T\geq N$, and note that $V_N(\cdot)\geq 0$ to give
	\begin{equation}\label{eq:T_costdec}
		\mathcal{J}_T(s) \leq V_N\Big(s_{0},K_N(s_0,\tilde{\mathbf{u}}_0)\Big) + \gamma^{N+R}\delta\sum_{j=0}^{T-1}\gamma^{i}
	\end{equation}
	By definition of $K_N(\cdot)$, we have
	\begin{equation*}
		V_N\Big(s_{0},K_N(s_0,\tilde{\mathbf{u}}_0)\Big) \leq \sum_{j=0}^{N+R-1}\gamma^{j}c(\hat{s}_j,\hat{u}_j) + \gamma^{N+R}\hat{J}(\hat{s}_N)
	\end{equation*}
    We combine this inequality with~\eqref{eq:T_costdec} and subtract~$\mathcal{J}_T(s)$ to give~\eqref{eq:full_performance}.
\end{proof}

We now provide a few corollaries of Lemma~\ref{lem:performance} that better illustrate the effect of the horizon length~$N$, rollout length~$R$, and Bellman-error~$\delta$ on the transient and long-term performance of \ac{AC4MPC}.

\begin{corollary}\label{cor:trans_performance}
	If Assumption~\ref{as:critic} holds,~$\mathbb{S}$ is bounded, and~$\gamma\in(0,1)$, then there exists~$d>0$ such that
	\begin{equation}\label{eq:trans_performance}
		\mathcal{J}_T(s) - \hat{\mathcal{J}}_T(s)
		\leq \\ \gamma^{N+R}\left(d + \delta/(1-\gamma)\right)
	\end{equation}
	for all~$s\in\mathbb{S}$ and~$T\geq N+R$.
\end{corollary}

\begin{proof}
	Since~$\mathbb{S}$ is bounded and~$\hat{J}(s)$ is continuous, there exists~$d\geq 0$ such that~$\hat{J}(s)\leq d$ for all~$s\in\mathbb{S}$. Furthermore,~$c(\cdot)\geq 0$ and $\sum_{j=0}^{T-1}\gamma^j\leq 1/(1-\gamma)$. Apply these bounds to~\eqref{eq:full_performance} to give~\eqref{eq:trans_performance}
\end{proof}

The result in Corollary~\ref{cor:trans_performance} demonstrates that the transient performance of \ac{AC4MPC}, defined by~$\mathcal{J}_T(s)$, is no worse than the actor alone, defined by~$\hat{\mathcal{J}}_T(s)$, plus some constant controlled by the horizon length~$N$, rollout length~$R$, and the Bellman error~$\delta$. In particular, we note that longer horizon lengths and rollout lengths reduce this constant and thereby improve the performance guarantee. For the long-term performance of \ac{AC4MPC}, we have the following result.

\begin{corollary}[Long-term performance]
	\label{cor:lt_performance}
	If Assumption~\ref{as:critic} holds,~$\mathbb{S}$ is bounded, and~$\gamma\in(0,1)$, then
	\begin{equation}\label{eq:lt_performance}
		\limsup_{T\rightarrow\infty}\left(\mathcal{J}_T(s) - \hat{\mathcal{J}}_T(s)\right) \leq \gamma^{N+R}\left(\frac{2\delta}{1-\gamma}\right)
	\end{equation}
	for all~$s\in\mathbb{S}$.
\end{corollary}

\begin{proof}
	By repeated application of~\eqref{eq:bellman} to the system in~\eqref{eq:cl_pi}, we have that
	\begin{equation*}
		\hat{J}(s) - \hat{\mathcal{J}}_T(s) \leq \sum_{j=0}^{T-1}\gamma^j\delta + \gamma^T\hat{J}(\hat{s}_T)
	\end{equation*}
	We apply this inequality to~\eqref{eq:full_performance} to give
	\begin{equation*}
		\mathcal{J}_T(s) - \hat{\mathcal{J}}_T(s)
		\leq \\ \gamma^{N+R}\delta\left(\sum_{j=0}^{T-R-N-1}\gamma^j + \sum_{j=0}^{T-1}\!\!\gamma^{i}\right) + \gamma^T\hat{J}(\hat{s}_T)
	\end{equation*}
	By using upper bounds for geometric series, we have
	\begin{equation}\label{eq:performance_withJT}
		\mathcal{J}_T(s) - \hat{\mathcal{J}}_T(s)
		\leq \\ \gamma^{N+R}\left(\frac{2\delta}{1-\gamma}\right) + \gamma^T\hat{J}(\hat{s}_T)
	\end{equation}
	Since~$\mathbb{S}$ is bounded and~$\hat{J}(s)$ is continuous, there exists~$d\geq 0$ such that~$\hat{J}(\hat{s}_T)\leq d$. We take the~$\limsup$ of each side of~\eqref{eq:performance_withJT} as~$T\rightarrow\infty$ to give~\eqref{eq:lt_performance}.
\end{proof}

Corollary~\ref{cor:lt_performance} establishes that the long-term performance ($T\rightarrow\infty$) of \ac{AC4MPC} relative to the actor is upper bounded by a constant controlled by the horizon length~$N$, rollout length~$R$, and the Bellman error~$\delta$. As~$N+R$ increases or~$\delta$ decreases, this constant converges to zero. Thus, Corollary~\ref{cor:lt_performance} demonstrates that long horizon lengths~$N$ or rollout lengths~$R$ can compensate for a poor estimate of the value function~$\hat{J}(s)$ for a given policy~$\hat{\pi}(s)$. Conversely, an accurate estimate of the value function~$\hat{J}(s)$ for a given policy can allow for a short horizon~$N$ or rollout length~$R$ to be used in the \ac{AC4MPC} algorithm.

These observations are again consistent with results for~$\ell$-step lookahead algorithms in dynamic programming (see, e.g., \cite{Bertsekas2023}). We note, however, that dynamic programming typically assumes that a globally optimal solution is obtained for the~$\ell$-step lookahead minimization. Thus, longer horizon lengths, in fact, bring the closed-loop performance closer to the optimal closed-loop performance of the system. Since we are permitting suboptimal solutions in the \ac{AC4MPC} algorithm, the best guarantee we obtain is that the closed-loop performance is bounded by the closed-loop performance of the actor used in the \ac{AC4MPC} algorithm. In economic MPC, similar results are obtained with respect to a periodic reference trajectory that is used to construct the terminal cost and constraint\cite{angeli2011average,amrit2011economic}.

We emphasize that the bounds in~\eqref{eq:trans_performance} and~\eqref{eq:lt_performance} are conservative. In practice, we expect \ac{AC4MPC} with moderate horizon lengths~$N$, rollout lengths~$R$, and Bellman errors~$\delta$ to outperform the actor. Stronger guarantees may be possible if we strengthen the assumptions on the stage cost and system, e.g., strict dissipativity or turnpike properties \cite{muller2016economic}.

The differences between the horizon length~$N$ and rollout length~$R$ are not obvious from these theoretical results, as the parameters appear as a sum~$N+R$ in each of the bounds. In practice, however, these parameters have different purposes and effects in the \ac{AC4MPC} algorithm. Specifically, longer horizons~$N$ increase the number of (free) decision variables in the optimization problem. Thus, increasing~$N$ generally increases the computational cost of solving the optimization problem but also permits more performance improvements of the \ac{AC4MPC} algorithm relative to the actor~$\hat{\pi}(\cdot)$. Conversely, increasing the rollout length~$R$ does not increase the number of (free) decision variables but can still mitigate the effect of the Bellman error~$\delta$. Thus, we can offer a simple recommendation for adjusting these two parameters:
\begin{itemize}
    \item If the performance of the actor~$\hat{\pi}(\cdot)$ is poor ($\hat{\mathcal{J}}_T(\cdot)$ is large), then longer horizon lengths~$N$ should be used. These longer horizon lengths can significantly increase the performance of the \ac{AC4MPC} algorithm relative to this actor~$\hat{\pi}(\cdot)$ by allowing \ac{AC4MPC} more flexibility to find a superior policy. 
    \item If the value function estimate~$\hat{J}(\cdot)$ is poor ($\delta$ is large), then longer rollout lengths~$R$ should be used. These longer rollout lengths can mitigate performance loss in \ac{AC4MPC} due to this Bellman error with a smaller increase in computational cost relative to increasing the horizon length~$N$.
\end{itemize}

\section{Multiple Shooting and Real-Time Iterations for AC4MPC}
\label{sec:ac4mpc_parallel}
So far, \Ac{AC4MPC} was defined conceptually as a single shooting formulation without a practical algorithm to solve the \ac{MPC} problem~\eqref{eq:MPC_compact}. In the following, we propose a practical algorithm, namely AC4MPC-RTI, for which the results of Sect.~\ref{sec:ac4mpc_cost_reduction} apply, which significantly reduces the online computation time.

Particularly, we propose to use the \ac{RTI} scheme~\cite{Diehl2005} and multiple shooting~\cite{Bock1984}, which create additional challenges for the algorithm. Within the \ac{RTI} scheme and the multiple shooting formulation, the solution never converges. The local optimum is rather tracked over several time steps, cf., Sect.~\ref{sec:nmpc}. Therefore, an initial guess provided by a policy roll-out may only obtain a lower cost after several \ac{QP} steps.
In fact, the solution obtained after each \ac{QP} step may even be infeasible for the nonlinear system dynamics due to the multiple shooting formulation.
The cost of an infeasible trajectory, i.e., a trajectory with \emph{gaps}~$F(s,u)-s^+\neq 0$, is challenging to evaluate. 

\begin{figure}
	\centering
	\includegraphics[width=\linewidth,trim={5mm 0 12mm 0},clip]{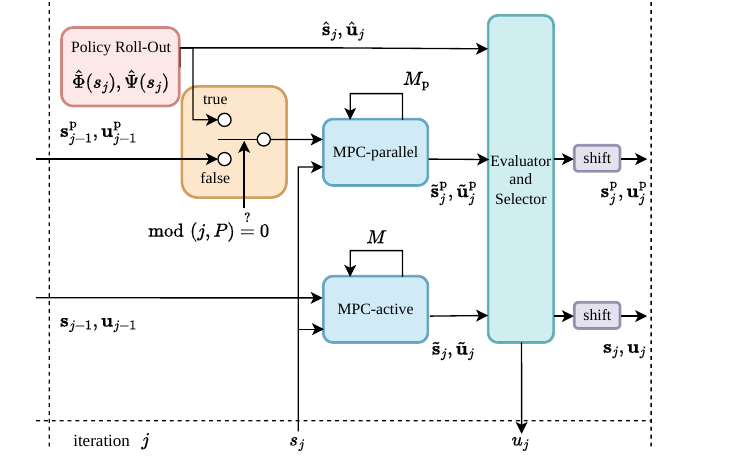}
	\caption{Algorithm sketch of \ac{AC4MPC}-RTI. In each iteration, the actor policy is rolled out to obtain a control and state trajectory (red). After each P iterations, the parallel MPC is initialized with the policy roll-out (yellow) and, otherwise, by the shifted previous MPC solution. The active MPC is initialized with the lowest-cost trajectory, which could either be the shifted solution of its last iteration, the parallel MPC trajectory, or the policy roll-out. The cost is provided by the proposed evaluation algorithm \emph{ac4eval} (green).}
	\label{fig:algorithm}
\end{figure}

To be compatible with the \ac{RTI} scheme, AC4MPC-RTI is extended by the following: 
(i) maintaining a state trajectory~$\mathbf{s}$ beside control trajectory~$\mathbf{u}$, 
(ii) \emph{allowing} trajectories to converge over multiple controller iterations by maintaining two MPC instances and reinitializing only one of them at all~$P$ iterations, 
(iii) adapting an evaluation algorithm that tackles the challenging cost prediction of an, usually infeasible, multiple shooting trajectory.
The basic algorithmic parts of AC4MPC-RTI are aligned with AC4MPC, see colored boxes in Alg.~\ref{alg:ac4mpc}, Alg.~\ref{alg:ac4mpc_rti} and Fig.~\ref{fig:algorithm}.

Addressing (i) and (ii), the parallel \ac{RTI} iteration scheme for AC4MPC-RTI with different initialization strategies is described in Sect.~\ref{sec:ac4mpc_parallelization} and referred to as parallelization. 
The evaluation algorithm related to (iii) is described in Sect.~\ref{sec:ac4mpc_eval}.
A schematic overview of~\ac{AC4MPC}-RTI is shown in Fig.~\ref{fig:algorithm}, and the algorithm in Alg.~\ref{alg:ac4mpc_rti}.

The proposed algorithm is generalizable to several parallel policy roll-outs, which could be obtained by differently trained \acp{NN}, cf., \emph{a mixture of experts}~\cite{Jacobs1991}. For clarity of exposition, we regard only one roll-out in the following.

\subsection{Parallelization}
\label{sec:ac4mpc_parallelization}
In AC4MPC-RTI, two \ac{MPC} instances and the policy roll-out are evaluated in each time step. 
In the \emph{parallel MPC}, the candidate trajectories~${\hat{\mathbf{s}}_j=\hat{\Phi}(s_j;\hat{\pi}(\cdot))}$ and~${\hat{\mathbf{u}}_j=\hat{\Psi}(s_j;\hat{\pi}(\cdot))}$ obtained from the policy roll-out are used as an initial guess for an~\ac{MPC}~\eqref{eq:MPC_general}, see~Fig.~\ref{fig:algorithm}. The~\ac{RTI} scheme then performs ~$M$ \ac{SQP} iterations starting from this initial guess. We use \ac{RTI} over several closed-loop time steps~$j$ to allow the solver to converge to the optimum over several closed-loop iterations. Particularly, the actor does not initialize the~\ac{MPC} solver in each iteration~$j$ but rather in all~$P\in\mathbb{N}^+$ time steps, where~$\mathrm{mod}(j,P)=0$.
If~$\mathrm{mod}(j,P)\neq0$, the initial guess for the parallel \ac{MPC} solver is obtained by \emph{shifting}.
The active solver uses the \ac{RTI} scheme with the previous shifted solution, where the previous solution is the lowest-cost solution of either the active solver, the parallel solver, or the pure actor rollout.
Shifting, as described previously for \ac{AC4MPC}, shifts the primal variable of the \ac{MPC} problem and simulates the system with the actor for the very last initial state, i.e., the controls are shifted by~$\zeta(s,\mathbf{u};\hat{\pi}(\cdot)$, and the states~$\mathbf{s}$ are shifted by
\begin{equation*}
	\xi(\mathbf{s};\hat{\pi}(\cdot)) := \big(s_1,\ldots,s_N,F(s_N,\hat{\pi}(s_N))\big).
\end{equation*}
Notably, in \ac{AC4MPC}-RTI, also the states~$\mathbf{s}$ are shifted and stored in each iteration and for both solvers as required for the RTI scheme~\cite{Diehl2005}.

If the trajectory~$(\hat{\mathbf{s}}_j,\hat{\mathbf{u}}_j)$ obtained from the policy roll-out or the parallel optimized trajectory~$(\tilde{{\mathbf{s}}}^\mathrm{p}_j,\tilde{\mathbf{u}}^\mathrm{p}_j)$ is superior in terms of the evaluated cost (see Sect.~\ref{sec:ac4mpc_eval}), the related states are used to initialize the active solver in the next iteration.
Given that the evaluated cost of the trajectory~$(\tilde{{\mathbf{s}}}_j,\tilde{\mathbf{u}}_j)$ obtained from the active solver is lowest, the active solver is not reinitialized with any policy, rather \acp{RTI}, or generally~$M$ \ac{SQP} iterations, are performed with successively starting at the shifted previous solution.
This guarantees that \ac{AC4MPC}-RTI performs at least as well as a \ac{MPC} formulation using~\ac{RTI}, yet with the computational burden of parallel policy evaluations.

\IncMargin{1em}
\begin{algorithm}
	\SetKwData{Left}{left}
	\SetKwData{This}{this}
	\SetKwData{Up}{up}
	\SetKwData{MPCact}{MPC-active}
	\SetKwData{MPCpar}{MPC-parallel}
	\SetKwData{Valfun}{Val.-Fun.}
	\SetKwData{IsLower}{IsLower}
	\SetKwFunction{Union}{Union}
	\SetKwFunction{FindCompress}{FindCompress}
	\SetKwInOut{Input}{input}
	\SetKwInOut{Output}{output}
	\Input{Policy~$\hat{\pi}(\cdot)$, value function~$\hat{Q}(\cdot)$ or~$\hat{J}(\cdot)$, max. \ac{SQP} iterations~$M,M_\mathrm{p}$, re-ini. period~$P$, correction par.~$\alpha$}
	\BlankLine
	\emph{\MPCact~$\leftarrow$ MPC~\eqref{eq:MPC_general} with~$V_f\gets\hat{J}$ or~$\hat{Q}$}\;
	\emph{\MPCpar~$\leftarrow$ MPC~\eqref{eq:MPC_general} with~$V_f\gets\hat{J}$ or~$\hat{Q}$}\;
	
	\For{$j\leftarrow 0$ \KwTo $\infty$}{
        \tikzmk{A}
		\emph{$s\leftarrow$state measurement}\;
		\emph{policy roll-out }$(\hat{\mathbf{s}}, \hat{\mathbf{u}}) \leftarrow \big(\hat{\Phi}(s;\hat{\pi}(\cdot)), \hat{\Psi}(s;\hat{\pi}(\cdot))\big)$\;\tikzmk{B}
        \boxit{col_rollout}
        \tikzmk{A}
		\If{$(j \mod P) == 0$}{
			$(\mathbf{s}^\mathrm{p}, \mathbf{u}^\mathrm{p}) \leftarrow(\hat{\mathbf{s}}, \hat{\mathbf{u}})$\;
			\If{$j == 0$}{
				$(\mathbf{s}, \mathbf{u}) \leftarrow(\hat{\mathbf{s}}, \hat{\mathbf{u}})$\;
		}}
		
		\emph{initialize \MPCact $\leftarrow (\mathbf{s}, \mathbf{u})  $}\;
		\emph{initialize \MPCpar $\leftarrow (\mathbf{s}^\mathrm{p}, \mathbf{u}^\mathrm{p}) $}\;\tikzmk{B}
        \boxit{col_ini}
		\tikzmk{A}
		\emph{$M$ \ac{SQP} iter. for \MPCact}\;
		\emph{$M_\mathrm{p}$ \ac{SQP} iter. for \MPCpar}\;
		
		\emph{obtain solution $ (\mathbf{s}, \mathbf{u}) \leftarrow$ \MPCact }\;
		\emph{obtain solution $ (\mathbf{s}^\mathrm{p}, \mathbf{u}^\mathrm{p})\leftarrow$ \MPCpar}\;\tikzmk{B}
        \boxit{col_solve}
		\tikzmk{A}
		\If{$  \eval(\mathbf{s}^\mathrm{p}, \mathbf{u}^\mathrm{p}) \leq \eval(\mathbf{s}, \mathbf{u})$}{
			$ (\mathbf{s}, \mathbf{u}) \leftarrow (\mathbf{s}^\mathrm{p}, \mathbf{u}^\mathrm{p})$
		}
		\If{$  \eval(\hat{\mathbf{s}},\hat{\mathbf{u}}) \leq \eval(\mathbf{s}, \mathbf{u})$}{
			$ (\mathbf{s}, \mathbf{u}) \leftarrow (\hat{\mathbf{s}},\hat{\mathbf{u}})$
		}\tikzmk{B}
        \boxit{col_eval}
		\emph{apply $ u \leftarrow \mathbf{u}[0]$ to the system}\;
		\tikzmk{A}\emph{shifting $ (\mathbf{s}, \mathbf{u}) \leftarrow
			\xi(\mathbf{s}; \hat{\pi}(\cdot)),\;\zeta(s,\mathbf{u}; \hat{\pi}(\cdot))$}\;
		\emph{shifting $ (\mathbf{s}^\mathrm{p}, \mathbf{u}^\mathrm{p}) \leftarrow
			\xi(\mathbf{s}^\mathrm{p}; \hat{\pi}(\cdot)),\;\zeta(s^\mathrm{p},\mathbf{u}^\mathrm{p}; \hat{\pi}(\cdot))$}\;\tikzmk{B}
        \boxit{col_shift}
	}
	\caption{AC4MPC-RTI}\label{alg:ac4mpc_rti}
\end{algorithm}\DecMargin{1em}

\subsection{Evaluation}
\label{sec:ac4mpc_eval}
After each iteration, the candidates~$(\hat{\mathbf{s}}_j,\hat{\mathbf{u}}_j)$ obtained from the policy roll-out, the parallel sequentially optimized roll-outs~$(\tilde{\mathbf{s}}^\mathrm{p}_j,\tilde{\mathbf{u}}^\mathrm{p}_j)$, and the trajectory of the active solver~$(\tilde{\mathbf{s}}_j,\tilde{\mathbf{u}}_j)$ are evaluated and ranked among their lowest predicted cost.
Evaluating the expected closed-loop cost of the optimization problem defined by~\eqref{eq:mpc_cost} solved by multiple-shooting and \acp{RTI} is non-trivial due to the following.

First, the problem can only be evaluated on a finite horizon.
To approximate the infinite horizon, the critic is used in the evaluator to approximate the infinite horizon cost, such as in the \ac{MPC} formulation~\eqref{eq:MPC_general}.

Secondly, evaluating the expected closed-loop cost of a multiple-shooting scheme using \acp{RTI} is challenging because the dynamics constraints might not be satisfied within the \ac{SQP} iterations, i.e. the trajectory exhibits gaps~\cite{Tenny2004}.

Within globalization strategies of optimization algorithms for multiple shooting formulations, these gaps are typically combined with the objective via a merit function in order to obtain a single evaluation criterion.
These merit functions need large exact penalties to \emph{outweigh} the other objectives~\cite{Nocedal1999}.
The merit function serves the purpose of \emph{closing the gaps} over iteration but is not suited to evaluate the expected closed-loop cost due to the rather arbitrary choice of weights, given they are large enough and lead to a numerically stable optimization algorithm.
Additionally, with open gaps, the trajectory is not dynamically feasible and, thus not suited for an evaluation.

A straight-forward method to obtain a feasible trajectory would involve using the controls~$\mathbf{u}$ to simulate the system~$F(\cdot)$ forward, starting from the current state~$s$.
Trivially, the trajectory would be feasible.
However, for unstable systems, the obtained state trajectory may differ vastly from the multiple-shooting trajectory~$\mathbf{s}$, thus, yielding a very high evaluation cost. Since in each time step the open-loop trajectory is recomputed based on the state feedback, the obtained closed-loop trajectory would be stabilized by the control law. Therefore, also simulating the control law in the evaluation for infeasible trajectories yields a better prediction of the cost.

In the following, a feasibility projection method is proposed to evaluate any trajectory~$(\mathbf{s},\mathbf{u})$ of length~$N$, that uses the actor policy as a correcting control law associated with open gaps.
The method involves a homotopy parameter~$\alpha\in [0,1]$ that scales the impact of the correction law.
We use an auxiliary control law
\begin{equation}
	\label{eq:auxlaw}
	\bar{s}_{k+1} = F(\bar s_k, \bar u_k), \quad
	\bar{u}_k = {u}_k + \alpha \big( \hat{\pi}(\bar{s}_k) - \hat{\pi}({s}_k) \big)
\end{equation}
to simulate the system forward to obtain the simulated controls~$\bar{\mathbf{u}}=[\bar{u}_0,\ldots,\bar{u}_{N-1}]$ and states~$\bar{\mathbf{s}}=[\bar{s}_0,\ldots,\bar{s}_{N} ]$.
A parameter of~$\alpha=0$ would correspond to an open-loop forward simulation without feedback.
Notably, the auxiliary state trajectory~$\tilde{\mathbf{s}}$ obtained from the control law defined in~\eqref{eq:auxlaw} would only differ from the SQP solution of the states~$\hat{\mathbf{s}}$, if the states~$\hat{\mathbf{s}}$ were infeasible w.r.t. the dynamics function.

Moreover, along the lines of~\cite{Bertsekas2005} and as discussed in the previous section, the value function is approximated by a roll-out of the actor policy for~$R$ steps at the final state~$\bar{s}_N$ to obtain~$\bar{s}_{N+1},\ldots,\bar{s}_{N+R}$ and~$\bar{u}_{N},\ldots,\bar{u}_{N+R-1}$ and the final critic value at~$\bar{s}_{N+R}$, c.f.,  Alg.~\ref{alg:eval}.

\IncMargin{1em}
\begin{algorithm}
	\SetKwInOut{Input}{input}
	\SetKwInOut{Output}{output}
	\SetKwInOut{Parameter}{parameter}
	\Input{Trajectory~${\mathbf{s}}\in \R^{n_s\times N}, {\mathbf{u}}\in \R^{n_s\times (N-1)}$ }
	\Parameter{Policy $\hat{\pi}(\cdot)$, value function $\hat{Q}(\cdot)$ or $\hat{J}(\cdot)$, correction parameter~$\alpha\in[0,1]$, evaluation roll-out length~$R$}
	\BlankLine
	\emph{Initialize cost $c_\mathrm{r}\gets 0$ }\;
	\emph{Initialize state, control $\bar{s}_0={s}_0, \bar{u}_0={u}$ }\;
	\For{$k\leftarrow 0$ \KwTo $N-1$}{
		\emph{get aux. control $\bar{u}_{k}={u}_k + \alpha \left( \hat{\pi}(\bar{s}_k) - \hat{\pi}({s}_k)\right)$}\;
		\emph{update cost $c_\mathrm{r}\gets c_\mathrm{r}+c(\bar{s}_k,\bar{u}_{k})$}\;
		\emph{simulate system $\bar{s}_{k+1}=f(\bar{s}_{k}, \bar{u}_k)$}\;
	}
	\For{$k\leftarrow N$ \KwTo $N+R$}{
		\emph{update cost $c_\mathrm{r}\gets c_\mathrm{r}+c(\bar{s}_k,\hat\pi{(\bar{s}_{k})})$}\;
		\emph{policy roll-out $\bar{s}_{k+1}=f(\bar{s}_{k}, \hat\pi{(\bar{s}_{k})})$}\;
	}
	\If{$\hat{J}(\cdot)$}{
		\emph{terminal cost $c_\mathrm{r} \gets c_\mathrm{r} + \hat{J}(\bar{s}_{N+R})$}\;
	}
	\Else{
		\emph{terminal cost $c_\mathrm{r} \gets c_\mathrm{r} + \hat{Q}(\bar{s}_{N+R}, \hat{\pi}(\bar{s}_{N+R}))$}\;
	}
	\Return{\emph{accumulated cost $c_\mathrm{r}$}}
	\caption{ac4eval$(\cdot)$}\label{alg:eval}
\end{algorithm}\DecMargin{1em}

\section{Experiments}
\label{sec:experiments}
In the following, the properties and the performance of the proposed algorithms are highlighted.
In Sect.~\ref{sec:illustrative_example}, the properties of \ac{AC4MPC} are illustrated on a low-dimensional example.
In Sect.~\ref{sec:car_example}, \Ac{AC4MPC}-RTI is evaluated in a more realistic scenario of time-optimally overtaking vehicles.
First, in Sect.~\ref{sec:numerical_issues_mpcnn}, we discuss some important implementation issues when using \acp{NN} within an \ac{MPC}.

\subsection{Using Neural Networks within MPC}
\label{sec:numerical_issues_mpcnn}
Although \ac{MPC} solvers, such as \texttt{acaods}~\cite{Verschueren2021}, are capable of solving nonlinear and nonconvex programs, the expected performance depends to a major extent on the local smoothness of the model.  \acp{NN} may contradict local smoothness requirements, e.g., \texttt{ReLU} networks are not even continuously differentiable. Therefore, the proposed \ac{AC4MPC} and \ac{AC4MPC}-RTI algorithms require smooth activation functions, such as~$\tanh$-activation functions, which we use in the following experiments. 

For \ac{AC4MPC}, the interior point algorithm~\texttt{ipopt}~\cite{Waechter2006} is used to solve the optimization problem to a local optimum. For the \ac{AC4MPC}-RTI algorithm, we use \ac{SQP} iterations with the \ac{RTI} scheme and Gauss-Newton Hessian approximations for the stage costs and the constraints due to their favorable numerical properties~\cite{Diehl2005}. For the terminal value function, which is a \ac{NN} in the proposed algorithm, we set the Hessian matrix in the QP sub-problems to a diagonal matrix with small entries and only compute first-order derivatives since this increased the numerical robustness in the performed experiments. In the \ac{AD} example, the nonlinearity of the critic was occasionally preventing the solver from converging. Therefore, the influence of the critic was diminished by multiplying it in the terminal value function by a factor~$0\leq\beta\leq1$.

The software framework used within the evaluations included the interior point solver \texttt{ipopt}~\cite{Waechter2006} and automatic differentiation framework \texttt{CasADi}~\cite{Andersson2019} for implementing \ac{AC4MPC}. For the \ac{RTI} solver we used  \texttt{acados}~\cite{Verschueren2021} and the learning framework~\texttt{L4CasADi}~\cite{Salzmann2023a, Salzmann2023b} to interface~\texttt{Pytorch} models. The actor and critic networks were trained using \texttt{stable-baselines-3}~\cite{stable-baselines3}.  \\

\subsection{An Illustrative Example}
\label{sec:illustrative_example}
\begin{figure}
	\centering
	\includegraphics[width=\linewidth,trim={2mm 2mm 2mm 1mm},clip]{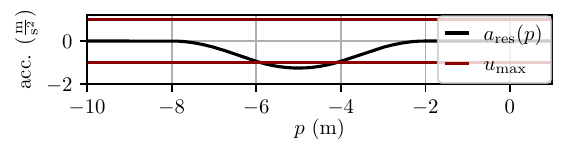}
	\caption{
		Force acting on the 1D vehicle due to a snowy slope and the maximum input acceleration force in the \snowhill{}.}
	\label{fig:winter_hill_forces}
\end{figure}

To shed light on the fundamental properties of \ac{AC4MPC}, an illustrative \snowhill{} is introduced. The environment models a point-mass vehicle with position~$p$ and velocity~$v$, with~$\dot{p}=v$ and the state~$s=[p,v]^\top$. The vehicle moves in one dimension and has to climb a \emph{snowy hill}, which is modeled by a force shown in Fig.~\ref{fig:winter_hill_forces}.
The force maximally decelerates at~$p=\text{-}5\,\mathrm{m}$ and is zero outside the interval~$p=[\text{-}8,\text{-}2]\mathrm{m}$.
The vehicle can be controlled by a bounded acceleration~$ |u| \leq 1\frac{\mathrm{m}}{\mathrm{s}^2}$, leading to the model equation~$\ddot{p}=\dot{v}=u+a_\mathrm{res}(p)$. The dynamics are discretized by an RK4 integrator and a discretization time of~$t_\mathrm{d}=0.1\mathrm{s}$ to yield the discrete-time system~$s_{k+1}=F(s_k,u_k)$. Notably, the control input is too low to directly drive the vehicle in certain states up the slope, i.e., in some positions, it has to move first away from the hill in order to catch enough speed to climb the slope.
Using the initial state~$\hat{s}_0$, the discrete-time \snowhill{} \ac{OCP}~\eqref{eq:MPC_illustrative_example_cont} is
\begin{align}
	\begin{split}
	\label{eq:MPC_illustrative_example_cont}
	&\min_{\substack{s_0,\ldots,s_{N_\mathrm{sim}},\\ u_0,\ldots,u_{N_\mathrm{sim}-1}}}\sum_{k=0}^{N_{\mathrm{sim}}} \sqrt{s_k^\top Q s_k+1} + \sum_{k=0}^{N_{\mathrm{sim}}-1}u_k^\top R u_k \\
	&\text{s.t. }
		s_0=\hat{s},\;  |u_k|\leq1,\; s_{k+1}=F(s_k,u_k), \; k\in \mathbb{N}_{N_\mathrm{sim}-1}.
	\end{split}
\end{align}

\begin{figure*}
	\centering
	\includegraphics[trim={2mm 2mm 2mm 2mm},clip,width=\linewidth]{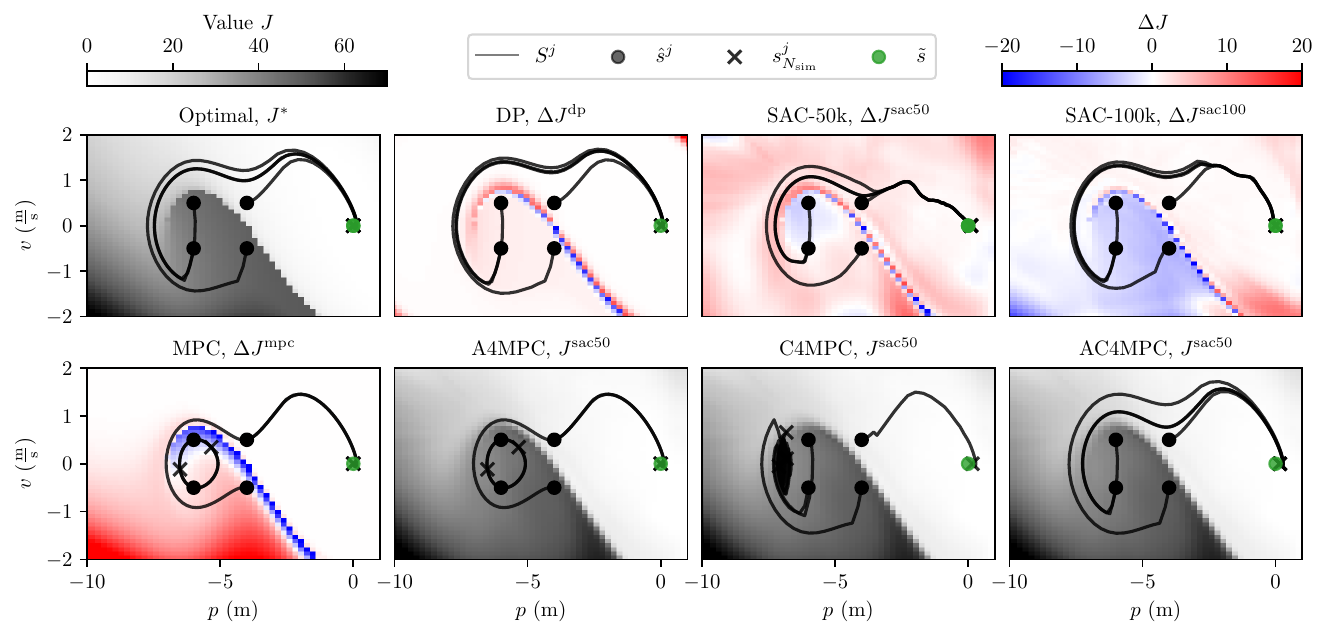}
	\caption{
	   Comparison of closed-loop trajectories~$S^j$ and relevant value functions for different control algorithms applied to the snow hill environment. Closed-loop trajectories are evaluated for four different starting states~$\hat{s}^j$, simulating the system for 20 seconds, following the related policy. The goal state~$\tilde{s}=[0,0]^\top$ can only be reached by certain algorithm variants. The first plots shows the ground truth value function~$J^*$ and trajectories obtained by~\ac{MPC} with a sufficiently long horizon to reach the goal state and constrained to the same. The next plot shows trajectories obtained by dynamic programming (DP), which are only close to optimal due to the discretization error. The difference~$\Delta J^\mathrm{dp}=J^*-J^\mathrm{dp}$ of the DP value function~$J^\mathrm{dp}$ to the optimal counterpart is shown. The next two plots show simulated trajectories using the actor, as well as the critic value function difference~$\Delta J^\mathrm{sac}$ to the optimal one of \ac{SAC} RL after~$5\cdot 10^4$ and~$10^5$ iterations, respectively. The lower left plot shows the nominal \ac{MPC} evaluation by initializing the trajectories at the current state and solving until full convergence. The value function~$J^\mathrm{mpc}$, corresponds to the open-loop values computed by the NMPC. Next, the A4MPC, which uses the actor obtained by \ac{SAC} to initialize each \ac{MPC} closed-loop iteration but no terminal value function, C4MPC which uses the initial state as initial guess the critic of the \ac{SAC} as terminal value function, and \ac{AC4MPC} which uses both, the actor and the critic of the \ac{SAC} are shown. The value functions plotted for A4MPC, C4MPC, and AC4MPC correspond to the \ac{SAC} critic value~$J^\mathrm{sac50}$, which is used directly in C4MPC and AC4MPC as the terminal value function, and the related policy roll-out is used in A4MPC and AC4MPC-RTI.}
	\label{fig:illustrative}
\end{figure*}

\begin{figure}
	\centering
	\includegraphics[trim={2mm 2mm 2mm 2mm},clip,width=\linewidth]{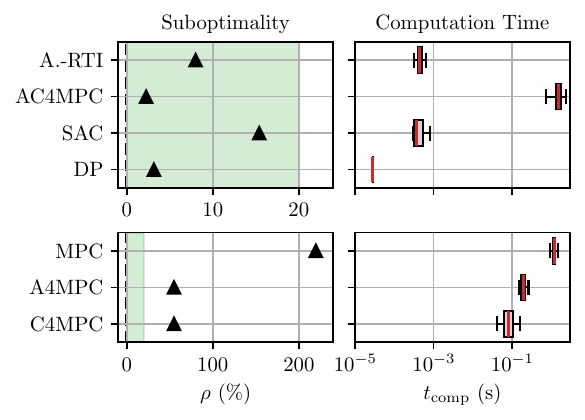}
	\caption{
	   Comparison of average suboptimality~$\rho=(J^{\{\cdot\}}-J^*)/J^*$ evaluated for closed-loop accumulated costs, corresponding to the closed-loop value functions, for different control algorithms on the \snowhill{}. The proposed \ac{AC4MPC} algorithm outperforms all other approaches, including the \ac{DP} that is slightly sub-optimal due to the discretization error of the state and control space. In this example, using only the critic (C4MPC) or the actor (A4MPC) leads to high costs, only slightly improving the nominal MPC. Using \ac{AC4MPC} has a high computational demand due to solving the optimization problem towards convergence in each iteration. Therefore, \ac{AC4MPC}-RTI significantly reduces the online computation time yet slightly increases the closed-loop cost. The cost of \ac{AC4MPC}-RTI is considerably lower than the \ac{RL} \ac{SAC} cost. The zoomed range in the upper plot is highlighted in green.}
	\label{fig:illustrative_performance}
\end{figure}

In the following, different control approaches for the \snowhill{} and related to \ac{AC4MPC} are compared qualitatively via samples of closed-loop trajectories and their value functions, cf., Fig.~\ref{fig:illustrative}.
Furthermore, a quantitative comparison of the obtained closed-loop cost for \ac{RL} variants, \ac{MPC}, \ac{AC4MPC} and \ac{AC4MPC}-RTI is given in Fig.~\ref{fig:illustrative_performance}.
For all experiments, we simulate for~$N_\mathrm{sim}=200$ steps.

First, the ``ground truth'' value function~$J^*$ and the policy are obtained by solving the \ac{OCP} as \ac{NLP} and fixing the final state to the goal state~$\tilde{s}=[0,0]^\top$, cf., top left plot in Fig.~\ref{fig:illustrative}. Distinct globally optimal trajectories are shown for different starting states~$\hat{s}$. Note that by fixing the final state and using an interior point solver~\texttt{ipopt}~\cite{Waechter2006}, the solver always converged.

Secondly, the value function~$J^\mathrm{dp}$ and policy are obtained by \ac{DP} within a discretization of~$\Delta s = [0.05\mathrm{m}, 0.05\frac{\mathrm{m}}{\mathrm{s}}]^\top$ and~$\Delta u = 0.01 \frac{\mathrm{m}}{\mathrm{s}^2}$, between~$v_\mathrm{eval}=[\shortminus3,3]\frac{\mathrm{m}}{\mathrm{s}}$ and~$p_\mathrm{eval}=[\shortminus12,4]\, \mathrm{m}$. 
Dynamic programming yields nearly optimal trajectories despite the state discretization error. In Fig.~\ref{fig:illustrative}, the difference to the optimal value function~$\Delta J^\mathrm{dp}=J^*-J^\mathrm{dp}$ is shown, in addition to example trajectories obtained by following the~\ac{DP} solution at each grid cell.

The policy obtained by \ac{SAC} after~$5\cdot10^4$ and~$10^5$ iterations and the critic function~$J^\mathrm{sac50}$ and~$J^\mathrm{sac100}$, respectively, are evaluated. For both the actor and the critic, feed-forward \acp{NN} with two layers of size 256 with~$\tanh$-activation functions are used. Notably, in \ac{SAC}, a Q-value function~$Q(s,u)$ is part of the algorithm. The regular value function is obtained by minimizing over the input~$u$ in each state. In Fig.~\ref{fig:illustrative}, it can be verified that the value function is approximated up to a small error, and the optimal policy drives the trajectories suboptimally to the goal state~$\tilde{s}$.

Thereafter, the nominal \ac{MPC} is evaluated using a terminal cost equal to the stage cost and a horizon of~$N_\mathrm{mpc}=20$.
The \ac{MPC} is initialized at the current state and solved with the \texttt{ipopt}~\cite{Waechter2006} solver towards convergence in each iteration. Fig.~\ref{fig:illustrative} reveals that \ac{MPC} gets occasionally stuck in local minima and can barely reach the goal state.
This is due to the missing terminal value function and initial guesses that lead to poor local minima.
Moreover, the horizon is too short to add a terminal constraint for the goal state.

\ac{AC4MPC} is evaluated, with two ablations. 
In the ablation named A4MPC, the actor is used to initialize the~\ac{MPC}. However, no terminal value function is used.
In the ablation C4MPC, the critic~$J^\mathrm{sac50}$ and~$J^\mathrm{sac100}$ are used as terminal value functions for the~\ac{MPC}.
Moreover, the current state and zero controls are used to initialize the primal variables of the \ac{MPC}.
In the latter three plots of~Fig.\ref{fig:illustrative} and in the performance comparison Fig.~\ref{fig:illustrative_performance}, it can be seen that only when using both the actor and the critic, ~\ac{AC4MPC} achieves superior performance.
In fact, in this example, the~\ac{AC4MPC} outperforms all other variants, including \ac{DP} in closed-loop performance.
The slightly worse performance of~\ac{DP} is due to discretization of the state space.

\begin{figure}
	\centering
	\includegraphics[trim={2mm 2mm 2mm 2mm},clip,width=\linewidth]{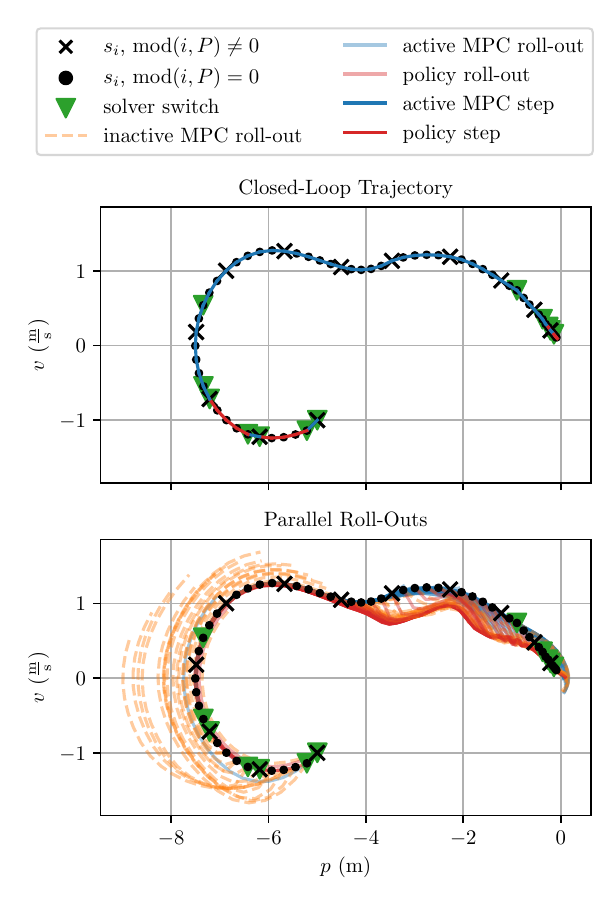}
	\caption{Phase plot of the AC4MPC-RTI closed-loop trajectory in the \snowhill{}, starting from the state~$s_0=[\shortminus5,\shortminus1]^\top$ and ending in the goal state~$\tilde{s}=[0,0]^\top$. At each~$P=5$ iterations, the parallel \ac{MPC} is re-initialized by using the actor policy. In each iteration, the actor is also rolled out by simulation. Using a cost evaluation parameter of~$\alpha=1$, the control corresponding to the lowest-cost trajectory is applied to the system. The upper plot shows whether an NMPC control was applied in the current time step (blue) or the proposed RL action (red). Additionally, green triangles indicate if, in the particular time step, the source of the output changed from either of the NMPC variants or the policy roll-out. The lower plot shows the parallel roll-outs of potentially both inactive NMPCs (orange) and the RL roll-out (red).}
	\label{fig:illustrative_traj}
\end{figure}

Finally, in Fig.~\ref{fig:illustrative_performance}, we quantitatively evaluate \ac{AC4MPC}-RTI for a horizon of~$20$ steps and actor and critic networks obtained after~$10^5$ or~$5\cdot 10^4$ \ac{SAC} steps, respectively. We used a correction parameter~$\alpha=1$ in Alg.~\ref{alg:eval} and an evaluation roll-out length~$R=20$. The results in Fig.~\ref{fig:illustrative_performance} highlight that \ac{AC4MPC}-RTI outperforms the corresponding \ac{SAC} variant. The computation time of \ac{AC4MPC}-RTI is over two orders of magnitude faster than~\ac{AC4MPC}, yet slower than the SAC policy evaluation. An illustrative single simulation of \ac{AC4MPC}-RTI, including open-loop planned trajectories, is shown for an initial state~$\hat{s}=[\shortminus5,\shortminus1]^\top$ in Fig.~\ref{fig:illustrative_traj}. It shows that the solver switches occasionally to the parallel MPC trajectory or the direct policy roll-out. The parallel MPC solver is reinitialized in all~$P=5$ steps.

In conclusion, the illustrative example highlights that, in general, both the actor and the critic approximations may be relevant for the \ac{AC4MPC} and that \ac{AC4MPC}-RTI significantly improves computation time by slightly trading-off performance. In the next section, a more elaborate example of \ac{AD} using \ac{AC4MPC}-RTI is given.

\subsection{Autonomous Driving}
\label{sec:car_example}
The following example considers a practically relevant and more involved scenario of autonomous driving.
The scenario includes a randomized road, i.e., a road that is constructed by randomizing its curvature~$\kappa(p_s)$ along the longitudinal position~$p_s$ in an interval~$[\ub{\kappa}, \lb{\kappa}]$.
Two slower \acp{SV} are simulated to follow a reference speed and a curvilinear path at random positions before a controlled \ac{EV}. All vehicles are simulated as five-state single-track models in the Frenet coordinate frame using ellipsoidal obstacle constraints, c.f.,~\cite{Reiter2023a}. The goal of the \ac{EV} is to overtake the~\acp{SV} while maintaining a speed limit~$\ub{v}=20\frac{\mathrm{m}}{\mathrm{s}}$, considering longitudinal and lateral acceleration constraints~$\ub{a}_\mathrm{lon}=3\frac{\mathrm{m}}{\mathrm{s}^2},\lb{a}_\mathrm{lon}=\shortminus12\frac{\mathrm{m}}{\mathrm{s}}$ and~$\ub{a}_\mathrm{lat}=5\frac{\mathrm{m}}{\mathrm{s}}$, respectively, and avoiding collisions. The single-track models are simulated by using parameters for the real-world vehicle \texttt{devbot 2.0} of the competition~\texttt{Roborace}~\cite{Roborace2020}. \\


As benchmark comparisons against the proposed \ac{AC4MPC}-RTI, a nominal \ac{MPC} that uses the \ac{RTI} scheme is implemented as in~\cite{Reiter2023}. Moreover, three \ac{RL} agents are trained by the \ac{SAC} method for~$2\cdot10^6$ steps or the \ac{PPO} method for~$10^7$ steps using different seeds for randomized initial \ac{NN} weights.
The nominal \ac{MPC} approximates time-optimal driving by avoiding the obstacles, yet without globalization strategy, i.e., the \ac{MPC} uses the \acp{RTI} purely based on the previous solution. It uses a zero-velocity terminal constraint. For \ac{AC4MPC}-RTI and the nominal \ac{MPC} prediction horizons~$N$ of~$10,30$ or~$60$ are used with a discretization time of~$t_\mathrm{d}=0.1\mathrm{s}$, a correction parameter~$\alpha=0$, a reinitialization parameter~$P=5$ and no evaluation roll-out, i.e.,~$R=0$.
Since in this example, the primal variables obtained during \ac{RTI} iterations exhibit only small open gaps, directly evaluate the multiple shooting trajectory cost, including penalties for open gaps. This cost can be easily obtained from numerical solvers, e.g., in \texttt{acados}~\cite{Verschueren2021}.
The \ac{SAC} and \ac{PPO} methods learn a critic and actor feed-forward \ac{NN} of two layers with~$256$ neurons each and smooth \texttt{tanh} activation functions.
The critic network of the \ac{PPO} method is a conventional value function of the state, whereas the critic network of the \ac{SAC} method is a Q-value function that also includes the control input~$u$. Therefore, the NMPC problem of AC4MPC-RTI based on the \ac{SAC} critic has two additional decision variables for~$u$ at the final stage.
The environment state used within this scenario consists of the ego vehicle state, curvature evaluations~$\kappa_i = \kappa(p_{\mathrm{s},i})$ with~$p_{\mathrm{s},i}=0,10,30,70,100,150$ and~$200$ meters lookahead distance of the current position, and the \ac{SV} states. In this example, the policy roll-out is not evaluated without optimizer iterations, i.e., lines~18 and 19 in Alg.~\ref{alg:ac4mpc_rti} do not apply.\\

\begin{figure}
	\centering
	\includegraphics[trim={2mm 2mm 2mm 2mm},clip,width=\linewidth]{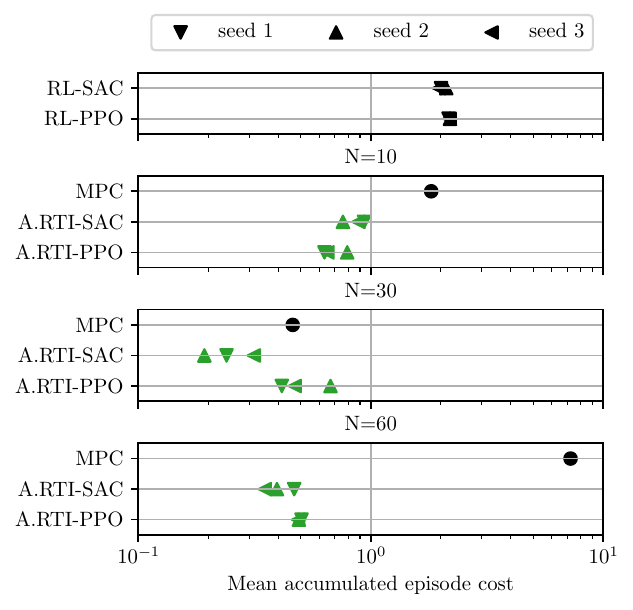}
	\caption{Accumulated mean episode cost of the \ac{AD} example with different prediction horizons~$N$ for various control algorithms. Three different training seeds were used for algorithms that include \acp{NN}. The nominal \ac{MPC} and \ac{RL} perform approximately equal at~$N=10$.}
	\label{fig:vehicle_performance}
\end{figure}

The algorithms are simulated in~$100$ random episodes with equal seeds among the approaches. The final closed-loop cost as defined within the \ac{MPC} and the \ac{RL} cost functions are summed for each episode and compared in Fig.~\ref{fig:vehicle_performance}.\\

The comparison reveals that the \ac{RL} policy performs similarly to the \ac{MPC} policy for a prediction horizon of~$N=10$. For a prediction horizon of~$N=30$, the \ac{MPC} outperforms the \ac{RL} agents significantly. For longer prediction horizons of~$N=60$, the \ac{MPC} gets occasionally stuck in local minima created by the obstacle and boundary constraints. This leads to a high closed-loop cost and a worse performance than the \ac{RL} agents, despite the higher computational demand, c.f.,~Tab.~\ref{tab:computation times}.
Tab.~\ref{tab:computation times} shows the averaged mean and maximum online solution time returned by the compiled \texttt{acados}~\cite{Verschueren2021} solver. For \ac{AC4MPC}-RTI, it computes the maximum computation time over all solvers, i.e., it assumes parallel processing and synchronization after each iteration. Notably, we do not account for other computation times, as these operations are assumed to be significantly faster than solving the optimization problem.

The proposed \ac{AC4MPC}-RTI algorithm outperforms both baseline approaches for short and longer horizons regarding closed-loop cost. For short horizons, the critic \ac{NN} provides a sufficient guess for the terminal value function, and the actor \ac{NN} is of minor importance. The main cost decrease for longer horizons stems from the actor \ac{NN} that helps escape from local optima. 
Notably, in this scenario, it was observed that the critic could also worsen the performance of the \ac{AC4MPC}-RTI approach.
In fact, the critic had to be scaled by a factor of~$0.1$. Otherwise, the MPC solver \texttt{acados}~\cite{Verschueren2021} did not converge sufficiently well.
This highlights the fact that \ac{AC4MPC}-RTI requires sufficiently well-trained and rather smooth \acp{NN} to achieve the proposed performance improvement.
However, \ac{AC4MPC}-RTI still achieved a superior performance with a higher computational burden as shown in Tab.~\ref{tab:computation times}. 
\begin{table}
	\centering
	\begin{tabular}{@{}l|ccc@{}}
		\addlinespace
		\toprule
		approach & \multicolumn{3}{c}{\emph{mean (maximum)} computation time in (ms)} \\
		& N=10 & N=30 & N=60  \\
		\midrule
		RL-SAC      			& &\multicolumn{1}{c}{0.37 (1.05)} & \\
		RL-PPO      		&	&\multicolumn{1}{c}{0.58 (2.70)}&\\
		MPC    			& 0.76 (1.74) 	& 2.53 (3.81) 	& 5.87 (10.15)\\
		AC4MPC-RTI (SAC)      			& 1.12 (1.88) 	& 3.35 (5.47) 	& 7.56 (13.78)\\
		AC4MPC-RTI (PPO)      			& 1.10 (2.86) 	& 3.12 (5.31) 	& 6.30 (8.89)\\
		\bottomrule
	\end{tabular}
	\caption{Online computation times (parallel evaluation) for \ac{AD} example.}
	\label{tab:computation times}
\end{table}

\begin{figure}
	\centering
	\includegraphics[scale=0.59,trim={0mm 9mm 0 3mm},clip]{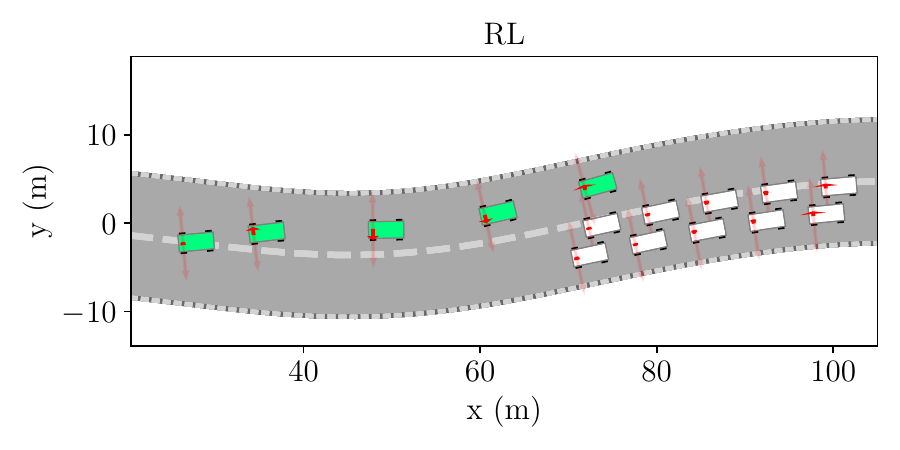}
	\includegraphics[scale=0.59,trim={0mm 9mm 0 3mm},clip]{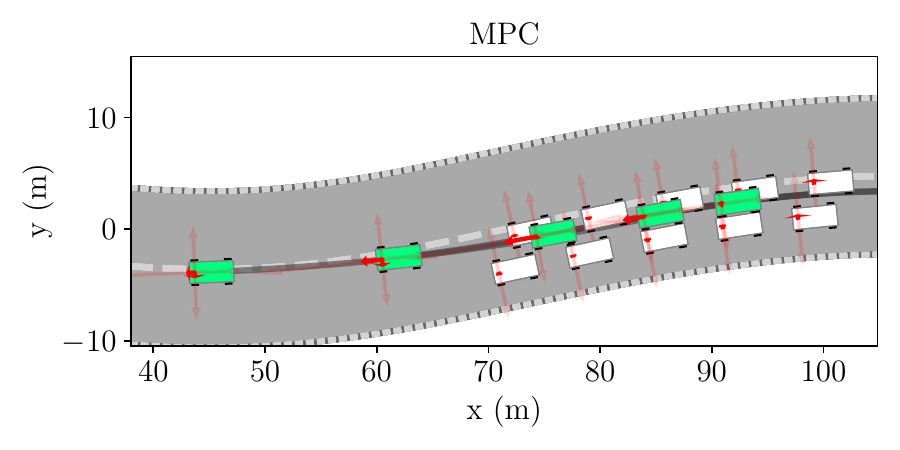}
	\includegraphics[scale=0.59,trim={0mm 1mm 0 3mm},clip]{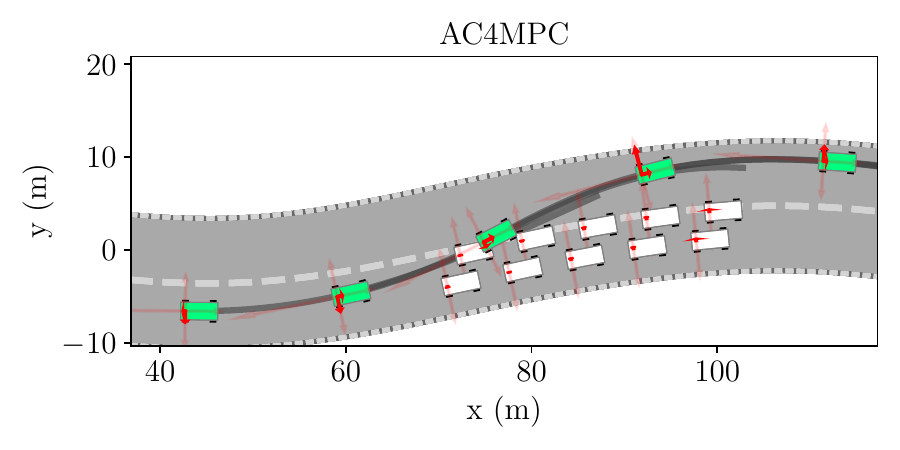}
	\caption{Snapshots at times~$t=6,7,8,9,10\mathrm{s}$ for an overtaking maneuver in a randomized scenario of the ego vehicle (green) of two surrounding vehicles (black) for the \ac{SAC} \ac{RL}, the \ac{MPC} and the \ac{AC4MPC}-RTI policies. The \ac{RL} policy lags behind \ac{MPC} and \ac{AC4MPC}, while the \ac{MPC} is stuck in a local minimum behind a leading vehicle.\ac{AC4MPC} escapes this minimum by a policy roll-out and swerves to the right. Red arrows indicate accelerations in the longitudinal and lateral directions in the vehicle coordinate frame. Planned trajectories are plotted in grey.}
	\label{fig:overtakiing_snapshots}
\end{figure}

Exemplary snapshots of the simulation are shown during the critical overtaking maneuver in Fig.~\ref{fig:overtakiing_snapshots}, and several rendered simulations can be seen at the website \url{https://rudolfreiter.github.io/ac4mpc_vis/}. 

The rendering of the simulation reveals that the RL agents progress conservatively and only overtake in the presence of larger gaps. As shown in Fig.~\ref{fig:overtakiing_snapshots}, MPC occasionally gets stuck behind vehicles due to the presence of local minima. \ac{AC4MPC}-RTI is able to escape this local minimum due to the critic in the terminal value function and the parallel policy roll-outs.

\acresetall
\section{Conclusion, Discussion, and Outlook}
\label{sec:discussion}
This work proposes a framework that can increase the performance of \ac{MPC} by using sufficiently well-trained \acp{NN} approximating the optimal policy and an optimal value function.  Training these networks is the main goal of \ac{RL}, and recently developed software tools, e.g.,~\cite{Salzmann2023b}, provide possibilities to merge these networks with \ac{MPC} solvers. Under some assumptions, we have shown the theoretical foundation of the proposed improvement in closed-loop performance. Practical, relevant examples provide experimental validation. Notably, the proposed algorithm can be easily parallelized to an ensemble of neural networks. 

In our particular experiments, the influence of the feasibility parameter~$\alpha$, c.f., Sect.~\ref{sec:ac4mpc_eval}, was small. We assume this is due to the \emph{minorly unstable} systems considered. In the \snowhill{} and autonomous driving example, the trajectory simulation within \ac{MPC} only minor gaps, leading to a minor influence of the feasibility parameter, only applies the actor control law for open gaps. However, in general, we expect an increased influence in highly unstable or chaotic systems.

The performance of the proposed algorithm depends on the quality of the trained \ac{RL} networks. However, this is trivially also true for the \ac{RL} policy. An ill-trained actor policy may not decrease the overall performance compared to conventional \ac{MPC}, assuming a long enough evaluation horizon. However, an ill-trained and, hence, highly nonlinear critic network used as a terminal value function may lead to numerical instabilities of the optimizer. In this case, the optimization algorithm may fail to converge. We observed such problems in the autonomous driving example of Sect.~\ref{sec:car_example} and mitigated it by scaling down the terminal value function by a weight of~$10$. Alternatively, the numerical properties of the value function can be adapted by either dedicated optimization problem-solving strategies or by enforcing favorable numerical properties already during the learning, such as in~\cite{Abdufattokhov2021,Karnchanachari2020}. Since the convergence problem related to the terminal value function was the main bottleneck of the proposed algorithm, this will be studied in future work.

\section{Acknowledgments}
The authors thank Johannes Köhler for his valuable feedback to this work.

\bibliographystyle{IEEEtran}
\bibliography{main}

\section{Biography Section}

\begin{IEEEbiography}[{\includegraphics[width=1in,height=1.25in,clip,keepaspectratio]{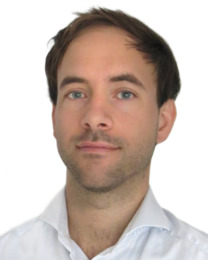}}]{Rudolf Reiter}{\space} received a master’s degree in electrical engineering from Graz University of Technology, Graz, Austria, in 2016, with a focus on control systems. He is 
currently pursuing a Ph.D. degree with the Marie-Skłodowska Curie Innovative Training Network position, University of Freiburg, Germany, under the supervision of Prof. Dr. Mortiz Diehl.
From 2016 to 2018, he worked as a Control Systems Specialist at the Anton Paar GmbH, Graz. From 2018 to 2021, he worked as a Researcher at the Virtual Vehicle Research Center, Graz. His research focuses on learning- and optimization-based motion planning and control for autonomous vehicles.
Mr. Reiter is an Active Member of the Autonomous Racing Graz Team.
\end{IEEEbiography}

\begin{IEEEbiography}[{\includegraphics[width=1in,height=1.25in,clip,keepaspectratio]{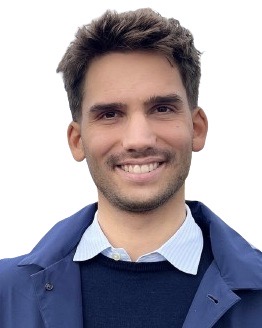}}]{Andrea Ghezzi}{\space} received his Master's degree cum laude in Automation and Control Engineering from Politecnico di Milano in 2020. After his graduation, he worked as a researcher in the Data Science R\&D Department at Tenaris in Dalmine, Italy. Since 2021, he has been a PhD student at the University of Freiburg under the supervision of Prof. Dr. Moritz Diehl. His research interests focus on numerical optimization and control, particularly on algorithms for mixed-integer nonlinear programming.
\end{IEEEbiography}

\begin{IEEEbiography}
[{\includegraphics[width=1in,height=1.25in,clip,keepaspectratio]{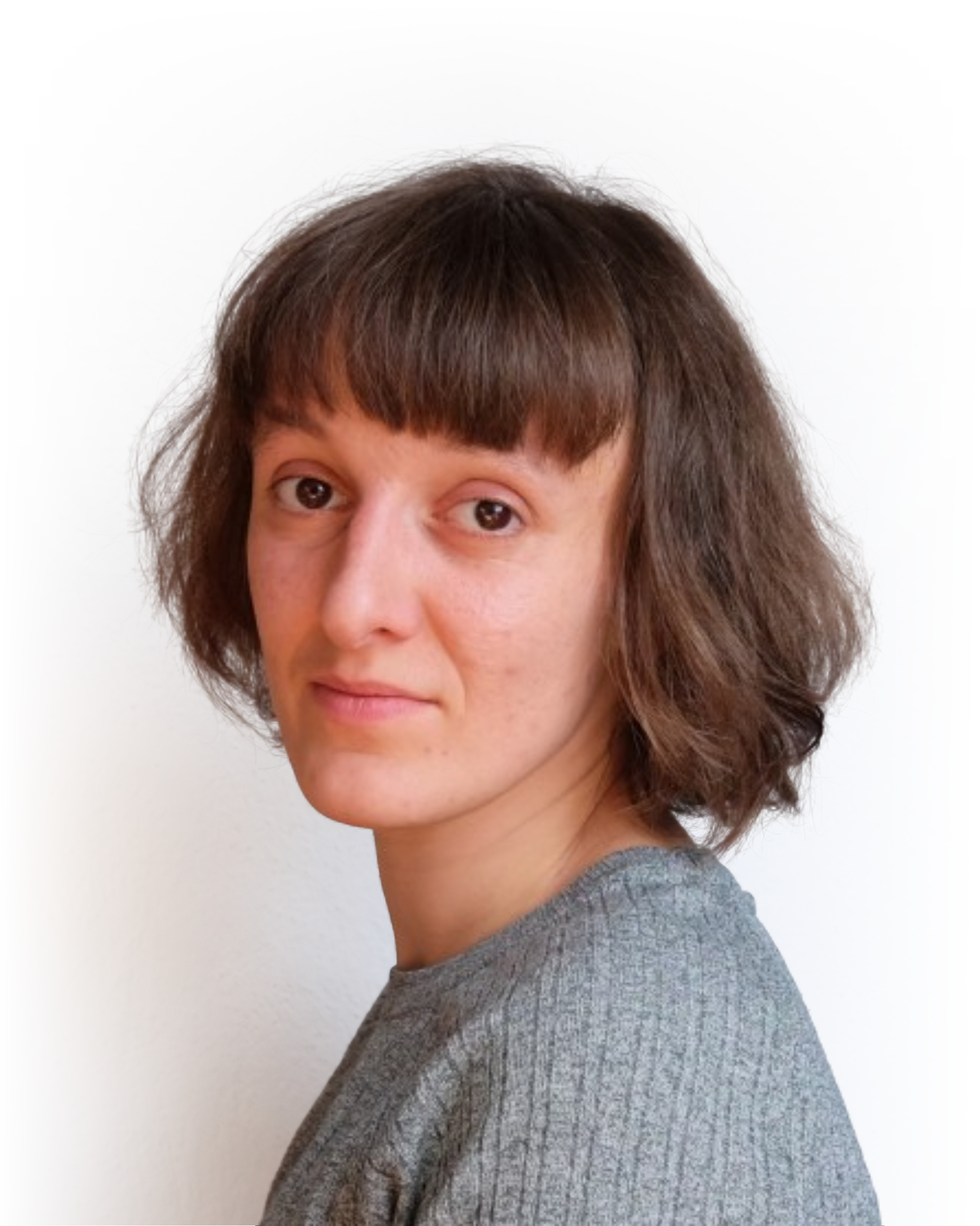}}]
{Katrin Baumgärtner}{\space} received her Master's degree in
computer science from the University of Freiburg,
Germany in 2019. Since 2020, she has been a
PhD student at the University of Freiburg under
the supervision of Prof. Dr. Moritz Diehl.
Her  research interests are structure-exploiting numerical methods for optimal feedback control and open-source software development.
\end{IEEEbiography}

\begin{IEEEbiography}[{\includegraphics[width=1in,height=1.25in,clip,keepaspectratio]{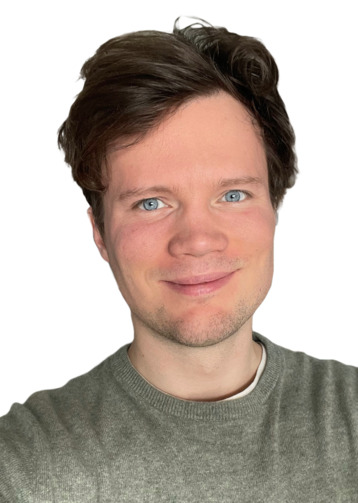}}]{Jasper Hoffmann}{\space} received his Master's degree in computer science from the University of Freiburg, Germany in 2020. Since 2021, he has been a PhD student at the University of Freiburg under the supervision of Prof. Dr. Joschka B\"{o}decker. His research interests focus on reinforcement learning and optimization-based control, particularly on rare events and combining learning and optimization-based control.
\end{IEEEbiography}

\begin{IEEEbiography}
[{\includegraphics[width=1in,height=1.25in,clip,keepaspectratio]{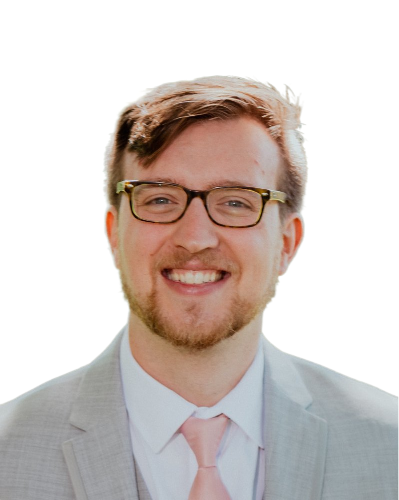}}]{Robert D. McAllister}
	Robert D. McAllister received the Bachelor of Chemical Engineering degree from the University of Delaware in 2017 and the Ph.D. degree in Chemical 
	Engineering from the University of California, Santa Barbara in 2022. He spent six months as postdoctoral research in the Delft Center for Systems and Controls (DCSC) at TU Delft. He is currently an assistant professor in the DCSC at TU Delft. His research interests include model predictive control, closed-loop scheduling, stochastic and distributional robustness of closed-loop systems, and data-driven control methods with applications in energy and agricultural systems. 
\end{IEEEbiography}

\begin{IEEEbiography}[{\includegraphics[width=1in,height=1.25in,clip,keepaspectratio]{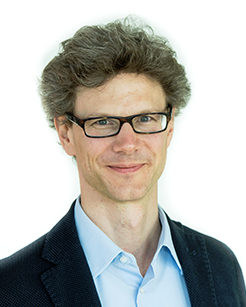}}]{Moritz Diehl}
	received his Diploma degree in mathematics and physics from Heidelberg University, Heidelberg, Germany, and Cambridge University, Cambridge, U.K., in 1999, and the Ph.D. degree in optimization and nonlinear model predictive control from the Interdisciplinary Center for Scientific Computing, Heidelberg University, in 2001. From 2006 to 2013, he was a Professor at the Department of Electrical Engineering at KU Leuven University Belgium and was the Principal Investigator with KU Leuven's Optimization in Engineering Center OPTEC. In 2013, he moved to the University of Freiburg, Germany, where he heads the Systems Control and Optimization Laboratory, Department of Microsystems Engineering (IMTEK), and is also with the Department of Mathematics. His research interests include optimization and control, spanning from numerical method development to applications in different branches of engineering, with a focus on embedded and on renewable energy systems.
\end{IEEEbiography}

\appendices
\section{Reinforcement Learning}
\label{sec:app_rl}
In the following, we describe two state-of-the-art actor-critic algorithms for off-policy and on-policy learning, namely \ac{SAC}~\cite{Haarnoja2018b} and \ac{PPO}~\cite{Schulman2017}.

\Ac{SAC} is an off-policy method using a replay buffer. It is based on the maximum-entropy \ac{RL} framework~\cite{Ziebart2008}, where an additional entropy term is added to the reward.
In \Ac{SAC}, the policy is stochastic during training, described by a parameterized Gaussian. The entropy term prevents the policy collapses to a single control, leading to better optimization and exploration properties during training~\cite{Ball2021}, and potentially leading to a more robust policy~\cite{Eysenbach2022}.
To derive a deterministic policy~$\hat{\pi}$ the maximum likelihood control is used.

We denote with~$\psi$ and~$\theta$ the parameters of the parameterized functions~$\hat{Q}_\psi$ and~$\hat{Q}_\theta$, e.g., the parameters of a neural network.
The additional entropy term introduced in maximum-entropy \ac{RL} changes the value function of~\eqref{eq:general_expected_return} to
\begin{align*}
	\begin{split}
		J^\textrm{soft}_\pi(s) \coloneqq& \E_\pi \left[ \sum_{k=0}^\infty \gamma^k \big[ c(s_k, u_k) + \alpha \; \mathcal{H}(\pi(\cdot \mid s_k)) \big] \right]\\
		&  s_0 = s,\; s_{k+1}= F(s_k,u_k),\; u_{k} \sim \pi(\cdot | s_k),
	\end{split}
\end{align*}
where~$\mathcal{H}$ denotes the entropy.
The influence of the entropy bonus can be controlled with the weight~$\alpha$.
Given the collected states in the replay buffer~$\mathcal{D}$, the policy objective of \Ac{SAC} is
\begin{equation*}
	\mathcal{L}_{\hat{\pi}}^\textrm{soft}(\psi) = \E_{s \sim \mathcal{D},\;  u \sim \hat{\pi}_\psi(s)}  \Big[ \hat{Q}_\theta (s, u) + \alpha \log(\hat{\pi}_\psi(u | s)) \Big].
\end{equation*}
The update of the critic is derived from the following loss
\begin{equation*}
	\mathcal{L}^\textrm{soft}_{\hat{Q}}(\theta) = \E_{s, u, s' \sim \mathcal{D}} \left[ \frac{1}{2} \big(c(s, u) + \gamma \hat{J}^\textrm{soft}_{\bar{\theta}} (s') \; - \; \hat{Q}_\theta(s, u) \big)^2 \right],
\end{equation*}
where the next state~$s'$ is derived from~$F(s, u)$ and the soft value function~$\hat{J}_\theta^\textrm{soft}$ is defined by 
\begin{equation*}
	\hat{J}^\textrm{soft}_{\bar{\theta}}(s) = \E_{u \sim \pi_\psi(s)} \Big[ \hat{Q}_{\bar{\theta}}(s, u) + \alpha \log \hat{\pi}(u | s) \Big].
\end{equation*}
The parameter~$\bar{\theta}$ indicates that it is a fixed copy of the parameter~$\theta$ that is periodically updated during training to stabilize the training~\cite{Mnih2015}.
For a detailed description of \Ac{SAC} we refer to~\cite{Haarnoja2018b}.

As a second actor-critic method, we consider \Ac{PPO}~\cite{Schulman2017}, an on-policy that collects multiple episodes of data before the policy and critic are updated.
As \Ac{PPO} is an on-policy method, transitions generated earlier in the training by outdated policies are neglected.
In practice, \Ac{PPO} is often used in combination with very fast simulation environments, where generating new samples comes with low computational time.
The main advantage of \Ac{PPO} is to prevent drastic updates that could destabilize the training by restricting the policy update via a simple clipping objective.
During training, a stochastic policy~${\hat{\pi}}_\psi$, i.e., often a parameterized Gaussian, is used. Assuming a given initial state~$s_0$, we draw a trajectory~$\tau$ by the forward simulation~$s_{k+1} = F(s_k, u_k)$ and~$u_k \sim \pi(s_k)$ until a maximum roll-out length~$M$.
The clipping objective of the critic~$\hat{J}_\theta$ is 
\begin{equation*}
	\mathcal{L}^\textrm{CLIP}_{\hat{J}} = \E_\tau \Big[ \frac{1}{2} \sum_{t=0}^{M - 1} \Big(c(s_k, u_k) +  \gamma \hat{J}_{\bar{\theta}} (s_{k+1}) \; - \; \hat{J}_\theta (s_k) \Big)^2 \Big].
\end{equation*}
To define the clipping objective for the policy, we require the probability ratio 
\begin{equation*}
	r_k(\psi) = \frac{\hat{\pi}_\psi(u_k|s_k)}{\hat{\pi}_{\bar{\psi}}(u_k|s_k)} \, ,
\end{equation*}
which measures how much the new policy~$\hat{\pi}_\psi$ changes with respect to the current policy~$\hat{\pi}_{\bar{\psi}}$ on the controls from the sampled trajectory~$\tau$.
Therefore, the \ac{PPO} clipping objective is then defined by
\begin{align*}
	\label{eq:ppo clip}
	\begin{split}
		\mathcal{L}_{\hat{\pi}}^\textrm{CLIP} (\psi) \coloneqq \E_\tau \Big[ \sum_{t=0}^{M - 1} &\min \Big\{ r_k (\psi) \hat{A}_\theta(s_k, u_k),\\
		&\textrm{clip}(r_k(\psi), 1- \epsilon, 1 + \epsilon) \; \hat{A}_\theta(s_k, u_k) \Big\} \Big]\,.
	\end{split}
\end{align*}
where~$\hat{A}_\theta(s, u)$ is the generalized advantage estimator~\cite{Schulman2016} derived from the learned value function~$\hat{J}_\theta$ by approximating the following target
\begin{align*}
	&\hat{A}_k \coloneqq \delta_k + (\gamma \lambda) \delta_{k+1} + \cdots + (\gamma \lambda)^{M - k + 1} \delta_{M - 1} \\
	&\textrm{with}\quad \delta_k \coloneqq c(s_k, u_k) + \gamma \hat{J}_\theta(s_{k+1}) - \hat{J}_\theta (s_k).
\end{align*}
Note that the $\textrm{clip}$ function projects the ratio~$r_l(\psi)$ to the interval from~$1 - \epsilon$ to~$1 + \epsilon$.
The parameter~$\lambda$ with~$0 \leq \lambda \leq 1$ trades-off bias and variance of the advantage estimator~$\hat{A}_k$.
For a detailed description of \ac{PPO}, we refer to \cite{Schulman2017}.

\end{document}